\documentclass[10pt]{iopart}

\usepackage[table]{xcolor} % color selected cells in table
\usepackage{graphicx}% Include figure files
\usepackage{dcolumn}% Align table columns on decimal point
\usepackage{bm}% bold math
\usepackage[mathlines]{lineno}% Enable numbering of text and display math
\usepackage{multirow}
\usepackage{color}
\usepackage{amssymb}
\usepackage[implicit=true,colorlinks=true,linkcolor=blue,
citecolor=blue,urlcolor=blue]{hyperref} % implicit: no eq,sec... as hyperref
\begin{document}
	
	\title[Manifestation of Majorana modes overlap in the Aharonov-Bohm effect]{Manifestation of Majorana modes overlap in the Aharonov-Bohm effect}
		
\author{S~V~Aksenov}

\address{%
	Kirensky Institute of Physics, Federal Research Center KSC SB RAS, Akademgorodok street 50/38, 660036 Krasnoyarsk, Russia}
\eads{\mailto{asv86@iph.krasn.ru}}
	
	\begin{abstract}

One of the key features of the Majorana bound states emerging in topological superconducting (SC) wires is increasing oscillations of their energy under the growth of magnetic field or chemical potential due to concomitant enhancement of hybridization of the Majorana mode wave functions initially localized at the opposite edges of the structure. In this study we found that the other consequence of it is a shift of Aharonov-Bohm (AB) oscillations of linear-response conductance in an interference device where two ends of the SC wire connected with a normal contact via non-SC wires (arms). In addition, it is accompanied by an oscillation period doubling. The numerical calculations for the spinful system are supported by the analytical results for different spinless models allowing to track the conductance evolution as the hybridization of the Majorana modes increases. It is shown that since the coupling between the different arms and normal contact is implemented only via the different-type Majoranas the AB oscillations acquire a fundamental $\pi/2$ shift in comparison with the effect for an analogous system of zero-energy quantum dots.

\end{abstract}	

%        71.10.Pm, % Fermions in reduced dimensions (anyons, composite fermions, Luttinger liquid, etc.)
%        74.45.+c, % Proximity effects; Andreev reflection; SN and SNS junctions
%        74.78.Na, % Mesoscopic and nanoscale systems
%        85.75.-d % Magnetoelectronics; spintronics: devices exploiting spin polarized transport or integrated magnetic fields
%\keywords{topological superconductivity, Majorana bound state, Andreev bound state, Aharonov-Bohm ring}
\maketitle
\ioptwocol

%\tableofcontents

\section{\label{sec1}Introduction}

Starting from theoretical works \cite{lutchyn-10,oreg-10} hybrid semiconducting/superconducting (SC) nanowires have become one of the most prospective systems where the Majorana bound states are being sought. It is proposed that in order to achieve an effective p-wave pairing, which is a key factor in the basic spinless model studied by Kitaev \cite{kitaev-01}, one has to combine a spin-orbit interaction, Zeeman splitting and s-type superconductivity \cite{stanescu-13b,sato-17}. First experiments on the local tunneling spectroscopy of InAs, InSb semiconducting nanowires possessing a large Rashba parameter and g-factor and being in contact with a superconductor revealed a zero-bias peak in conductance if the magnetic field exceeded some critical value \cite{mourik-12}. The effect was initially attributed to the emergence of zero-energy Majorana quasiparticle. 

The further progress in epitaxial growth of the hybrid nanowires allowed to synthesize the samples with uniform and clean interfaces between an SC shell and semiconducting core. It led to the hard SC gap induced by the proximity effect \cite{chang-15,pan-20} and allowed to approach the theory predicted $2G_{0}$-height ($G_{0}=e^2/h$ - the conductance quantum) of the zero-bias peak \cite{law-09,nichele-17,yu-21}. In spite of these advantages the nearly quantized peaks measured in the experiments are not sufficiently stable upon the variation of gate voltage and magnetic field \cite{nichele-17,zhang-21} and can be related to the presence of trivial Andreev bound states which are rather ubiquitous in the inhomogeneous wires \cite{prada-20,marra-22}. Moreover, other transport peculiarities related to the Majorana excitation have not been observed in practice yet. For example, the linear-response conductance resonances of two contacts coupled to the opposite ends of the wire have to be correlated provided by the nonlocality of Majorana bound state \cite{yu-21}. The measurement of nonlocal conductance should represent the features of the bulk-gap closing and reopening \cite{rosdahl-18,hess-21}. 

If the lengths of Majorana-state localization (inversely proportional to the induced gap) and SC part of the wire are comparable quantities the Zeeman-field dependence of conductance should demonstrate increasing oscillations when the strength of magnetic field grows up. Such a behavior takes place due to the overlap of wave functions of the Majorana modes (hereinafter, it is assumed that the Majorana bound state consists of two Majorana modes) \cite{dassarma-12,albrecht-16}. The experiment \cite{albrecht-16} revealed that the oscillations decay or remain pinned at zero energy. 

Many mechanisms were suggested to explain this discrepancy between the theory and experiment. The absence of oscillations can be caused by the Coulomb interaction \cite{dominguez-17} or inverse proximity effect due to drain lead coupled to the SC material \cite{danon-17}. The decaying oscillations, anticrossings or monotonous decay can arise because of magnetic-field orbital effects \cite{dmytruk-18}, nonuniform pairing potential \cite{fleckenstein-18}, nonuniform Rashba field \cite{cao-19} or the combined influence of temperature, multiple subbands and the simultaneous presence of Majorana and Andreev bound states in the SC wire \cite{chiu-17}.     

Different interference setups have been proposed to detect the nontrivial phase in the SC nanowire. If it is embedded in one of the arms of Aharonov-Bohm (AB) ring, then the Majorana excitation permits to transfer individual electrons and holes instead of Cooper pairs resulting in the change of the magnetoconductance oscillation period \cite{benjamin-10,akhmerov-11,whiticar-20} and generation of the $h/e$ harmonic of flux-independent persistent current \cite{buttiker-86,jacquod-13}. The ways to distinguish between the Majorana and Andreev bound states were analyzed as well \cite{tripathi-16,hell-18,whiticar-20}. Some studies were focused on the features of nonlocal transport through a quantum dot, where the topological SC wire plays the role of a structure side-attached to the main channel by both ends, i.e. having a $\Pi$-shape. \cite{liu-11,ramos-andrade-18}.

\begin{figure}[tb]
	\includegraphics[width=0.45\textwidth]{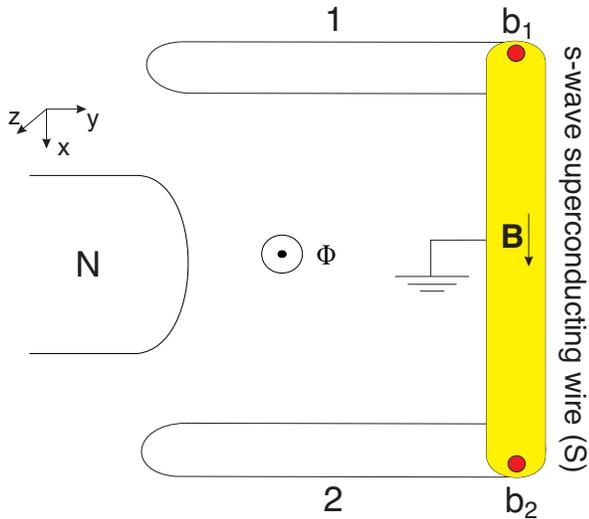}
	\caption{\label{fig1} Sketch of an $\Pi$-shape interference device. It is partly covered by an s-wave superconductor. The sections left uncovered form top (1) and bottom (2) arms which connect the two ends of the directly-grounded superconducting segment (S) with a massive normal contact (N). A magnetic field, $\mathbf{B}$, is applied locally parallel to the superconducting wire controlling its topological phase. In the nontrivial phase two Majorana modes, $b_{1}$ and $b_{2}$, emerge at the opposite edges of the wire as indicated by small red circles. A magnetic flux, $\Phi$, piercing the device plane induces the Aharonov-Bohm oscillations of conductance of the normal contact.}
		%Sketch of an interference device where arms connect two ends of an s-wave superconducting wire with a normal contact. A magnetic field $\mathbf{B}$ is applied locally parallel to the superconducting wire controlling its topological phase. In the nontrivial phase two Majorana modes, $b_{1}$ and $b_{2}$, emerge at the opposite edges of the wire as indicated by small red circles. A magnetic flux, $\Phi$, piercing the device plane induces the Aharonov-Bohm oscillations of conductance of the normal contact.}
\end{figure}

Here we propose the other way to test the overlap of the Majorana wave functions utilizing interference transport scheme. The setup sketched in Fig.\ref{fig1} includes a massive normal contact coupled with a $\Pi$-shape semiconducting wire partly covered by an SC material. The magnetic field is applied to this directly-grounded SC area allowing to switch between the trivial and nontrivial phases. Nonlocal transport properties of the similar devices (that implies two normal contacts) have been already investigated \cite{shang-14,wang-14,chi-21}. However, the AB effect and its differences for nonoverlapping and overlapping Majorana modes, $b_{1,2}$, (NOMMs and OMMs, respectively) were not considered. This study is devoted to these issues.

The rest of article is organized as follows. In Sec.\ref{sec2.1} we describe a spinful model of the interference device and present a theory of quantum transport based on the nonequilibrium Green's functions (GFs). Sec. \ref{sec2.2} is devoted to the numerical analysis of transport properties of the spinful model in the linear-response regime. In Sec.\ref{sec3.1} an analytical expression of conductance for a spinless model of the interference device is obtained. In Secs.\ref{sec3.2}--\ref{sec3.4} we discuss the AB effect differences in the situations of nonoverlapping, weakly hybridizing and overlapping Majorana modes, respectively. We conclude in Sec.\ref{sec4} with a summary. In \ref{apxA} equations of motions (EOMs) and retarded GFs for different low-energy models of the spinless interference device are provided. In \ref{apxB} transport properties of an interferometer containing zero-energy quantum dots (QDs) are numerically studied.

\section{\label{sec2} Spinful interference system}

\subsection{\label{sec2.1} Hamiltonian and transport equations}

One can treat the system under consideration as the normal (N) and superconducting (S) contacts coupled by the top (1) and bottom (2) arms, which are one-dimensional normal wires, as it is schematically depicted in Fig. \ref{fig1}. The model Hamiltonian has a following general form:
\begin{equation} \label{H}
\hat{H}=\hat{H}_{N}+\hat{H}_{D}+\hat{H}_{t}.
\end{equation}
The summand $\hat{H}_{N}$ characterizes the single-band normal contact,
\begin{equation} \label{HN}
\hat{H}_{N} =\sum\limits_{k\sigma}\left(\xi_{k}-\frac{eV}{2}\right)c^+_{k\sigma}c_{k\sigma},
\end{equation}
where $c_{k\sigma}$ - an annihilation operator of the electron with momentum $k$, spin $\sigma$ and energy $\xi_{k}=\varepsilon_{k}-\mu$ in the contact; $\mu$ - a chemical potential of the system; $eV$ - an electric-field energy related to bias voltage. The second term is a tight-binding Hamiltonian which models a device consisting of two normal wires (or top and bottom arms) coupled to the opposite ends of the SC wire, $\hat{H}_{D}=\hat{H}_{1}+\hat{H}_{2}+\hat{H}_{S}+\hat{H}_{\tau}$, 
\begin{eqnarray} 
&&\hat{H}_{i} =\xi_{i}\sum\limits_{\sigma;j=1}^{N_{i}}d^+_{ij\sigma}d_{ij\sigma}-\label{HA}\\
&&~~~~~~-\sum\limits_{j=1}^{N_{i}-1}\left(\frac{t_{i}}{2}d_{ij\sigma}^{+}d_{i,j+1,\sigma}+i\frac{\alpha^{x}_{i}}{2}d_{ij\sigma}^{+}d_{i,j+1,\bar{\sigma}}+H.c.\right),\nonumber\\
&&\hat{H}_{S} =\sum\limits_{j=1}^{N}\left[\xi_{S}\sum\limits_{\sigma}a^+_{j\sigma}a_{j\sigma}-ha^+_{j\uparrow}a_{j\downarrow}+
\Delta \left(a_{j\uparrow}a_{j\downarrow} + H.c.\right)\right]-\nonumber\\
&&~~~~~~~-\sum\limits_{\sigma;j=1}^{N-1}\left[\frac{t}{2}a^+_{j\sigma}a_{j+1,\sigma}+\frac{\alpha^{y}}{2}\sigma a^+_{j\sigma}a_{j+1,\overline{\sigma}}+H.c.\right],\label{HS}\\
&&\hat{H}_{\tau}=-\sum\limits_{\sigma}\left(\frac{\tau_{1}}{2}d_{1,N_{1},\sigma}^{+}a_{1\sigma}+\frac{\tau_{2}}{2}d_{21\sigma}^{+}a_{N\sigma}+\right.\label{Htau} \\
&&\left.~~~~~~~~~~~~~~~~+ i\frac{\alpha^{x}_{i}}{2}d_{1,N_{1},\sigma}^{+}a_{1\bar{\sigma}}-i\frac{\alpha^{x}_{i}}{2}d_{21\sigma}^{+}a_{N\bar{\sigma}}+H.c.\right),\nonumber
\end{eqnarray}
where $\xi_{i}=\varepsilon_{i}-\mu,~t_{i},~\alpha_{i}^{x}$ - a single-electron energy, hopping parameter and intensity of the Rashba spin-orbit interaction in the $i$th arm ($i=1,2$), respectively; $\xi_{S}=\varepsilon-\mu,~t,~\alpha^{y}$ - the analogous quantities in the SC segment of the device; $\Delta$ - a strength of SC pairing induced by the proximity effect; $\tau_{i}$ - a parameter of hopping between the $i$th arm and SC wire; $h$ - the Zeeman splitting energy caused by the longitudinal magnetic field, $\mathbf{B}$ (see Fig. \ref{fig1}). Note that to realize the topologically nontrivial phase and the Majorana modes, $b_{1}$ and $b_{2}$, the magnetic field has to be oriented perpendicular to the vector of effective Rashba field in the SC segment \cite{lutchyn-10,oreg-10}. The SC wire and $i$th arm contains $N$ and $N_{i}$ sites, respectively.

The last term in (\ref{H}) takes into account tunneling processes between the normal contact and structure, 
\begin{equation} \label{Ht}
\hat{H}_{t} =\sum\limits_{k\sigma}c_{k\sigma}^{+}\left(t_{1}e^{-i\frac{\phi}{2}}d_{11\sigma}+
t_{2}e^{i\frac{\phi}{2}}d_{2,N_{2},\sigma}\right) + H.c.,
\end{equation}
where $t_{1,2}$ - intensities of tunnel couplings; $\phi=2\pi\Phi/\Phi_{0}$ - the AB phase induced by a magnetic flux $\Phi$ penetrating the device; $\Phi_{0}=h/e$ - the flux quantum.

To calculate a stationary charge current in the normal contact, $I=e\frac{\partial }{\partial t}\sum_{k\sigma}\langle c_{k\sigma}^{+}(t)c_{k\sigma}(t)\rangle$, the Hamiltonian (\ref{H}) is diagonalized based on the Gorkov-Nambu operators \cite{gorkov-58,nambu-60}, $\hat{f}_{l}=\left(f_{l\uparrow},~f_{l\downarrow}^{+},~ f_{l\downarrow},~f_{l\uparrow}^{+}\right)^T$, where $f_{l\sigma}$ - the second quantization operator of particle in one of the four above-described subsystems, $l$ - an index of site or wave vector. Matrix nonequilibrium GFs \cite{keldysh-65} for the introduced field operators can be defined as  
$\hat{G}_{lm}^{ab}\left(\tau_{a},\tau_{b}\right)=-i
\left\langle T_{K} \hat{f}_{l}(\tau_{a}) \hat{f}_{m}^+(\tau_{b}) \right\rangle$, where $T_{K}$ - an operator of time ordering on the Keldysh contour. The indices $a,b=+,-$ of the GFs indicate that the time argument of the corresponding operator belongs to either lower or upper branch of the Keldysh contour.

Performing a series of transformations, the following expression for the current is obtained \cite{valkov-19,aksenov-20}:
\begin{eqnarray} \label{IL3}
&&I=
\frac{2e}{h}\sum_{i,j=1,2}\int\limits_{-\infty}^{+\infty}d\omega
Tr\Biggl[\hat{\sigma}Re\Biggl\{
\hat{\Sigma}_{ij}^{r}\left(\omega\right)\hat{G}_{ji}^{+-}\left(\omega\right)+\Biggr.\Biggr.\\
&&\Biggl.\Biggl.~~~~~~~~~~~~~~~~~~~~~~~~~~~~~~~~~~~~~~~~~+\hat{\Sigma}_{ij}^{+-}\left(\omega\right)
\hat{G}_{ji}^{a}\left(\omega\right) \Biggr\}\Biggr],\nonumber
\end{eqnarray}
where $\hat{\sigma}=diag\left(1,-1,1,-1\right)$. The matrix blocks of the device's advanced GF, $\hat{G}_{ji}^{a}$, are determined by the solution of the Dyson equation,
\begin{equation}\label{Ga}
\hat{G}^{a}=\left[\left(\omega-\hat{h}_{D}-\hat{\Sigma}^{r}\left(\omega\right)\right)^{-1}\right]^{+},
\end{equation}
where $\hat{\Sigma}^{r}\left(\omega\right)$ - a matrix of the retarded self-energy function reflecting the influence of the contact on the device. Assuming that the normal contact band is wide, one has the following nonzero blocks of $\hat{\Sigma}^{r}\left(\omega\right)$:
\begin{eqnarray}\label{Sr}
&&\hat{\Sigma}^{r}_{11\left(22\right)}=-\frac{i}{2}\hat{\Gamma}_{11\left(22\right)},~\hat{\Sigma}^{r}_{21\left(12\right)}=-\frac{i}{2}\hat{\Gamma}_{12}\hat{\Phi}_{\pm}^{2},\\
&&\hat{\Phi}_{\pm}=diag\left(e^{ \pm i\frac{\phi}{2}},e^{\mp i\frac{\phi}{2}},e^{ \pm i\frac{\phi}{2}},e^{\mp i\frac{\phi}{2}}\right),\nonumber
\end{eqnarray}
where $\hat{\Gamma}_{ii}=\Gamma_{i}\hat{\tau}_{0}$, $\Gamma_{i}=2\pi\rho\left(t_{i}/2\right)^2$ - functions characterizing the broadening of the device levels ($i=1,2$); $\hat{\tau}_{0}$ - an $4\times4$ identity matrix; $\rho$ - the contact density of states; $\Gamma_{12}=\sqrt{\Gamma_{1}\Gamma_{2}}$. The matrix lesser GF, $\hat{G}_{ij}^{+-}$, in (\ref{IL3}) is found from the solution of the Keldysh equation, $\hat{G}^{+-}=\hat{G}^{r}\hat{\Sigma}^{+-}\hat{G}^{a}$. Its nonzero blocks are 
\begin{equation}\label{SpmSr}
\hat{\Sigma}_{ij}^{+-}=-2\hat{\Sigma}_{ij}^{r}\cdot diag\left(n_{h},~n_{e},~n_{h},~n_{e}\right),~i,j=1,2,
\end{equation}
where $n_{e\left(h\right)}\equiv n\left(\omega\mp eV/2\right)$ - the Fermi-Dirac distribution functions.

\subsection{\label{sec2.2} Results of numerical calculation}

For simplicity, in the following numerical analysis we will assume that $\xi_{1}=\xi_{2}$, $t_{1}=t_{2}=t$, $N_{1}=N_{2}$. At the same time, to study the influence of spin-orbit coupling in the arms on the AB effect in more details the cases of identical and different normal wires will be analyzed separately, namely: 1) $\alpha_{1}^{x}=-\alpha_{2}^{x}=\alpha^{x}$; 2) $\alpha_{1}^{x}=-2\alpha_{2}^{x}=\alpha^{x}$. In order to model real InAs, InSb semiconducting wires of $\mu m$ length the lattice constant is chosen as $a=50$ nm. Then, using the effective mass value found experimentally, $m=0.015m_{0}$, the hopping parameter is $t=\hbar^2/ma^2\approx 2~meV$ \cite{mourik-12}. Throughout all the numerical calculations the other energy parameters are measured in the units of $t$. Namely, $\varepsilon_{1,2}=0$, $\varepsilon=1$, $\mu=0$, $\Delta=0.2$, $\alpha^{x}=0.4$, $\alpha^{y}=0.3$, $\Gamma_{1,2}=0.01$, $\tau_{1,2}=0.1$, $k_{B}T\approx0$, $N_{1}=30$ and $N=60$.

\begin{figure}[tb]
	\includegraphics[width=0.525\textwidth]{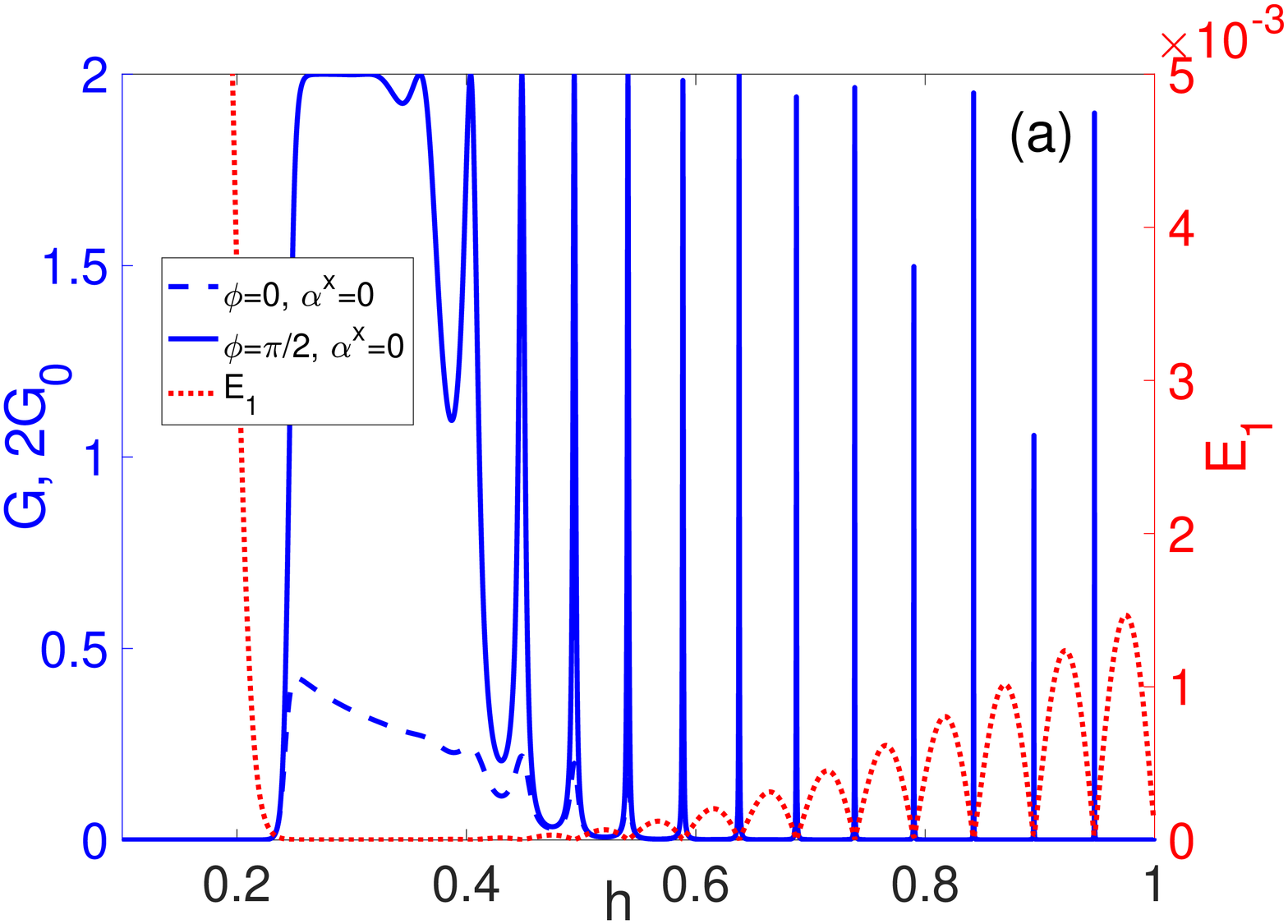}
	\includegraphics[width=0.5\textwidth]{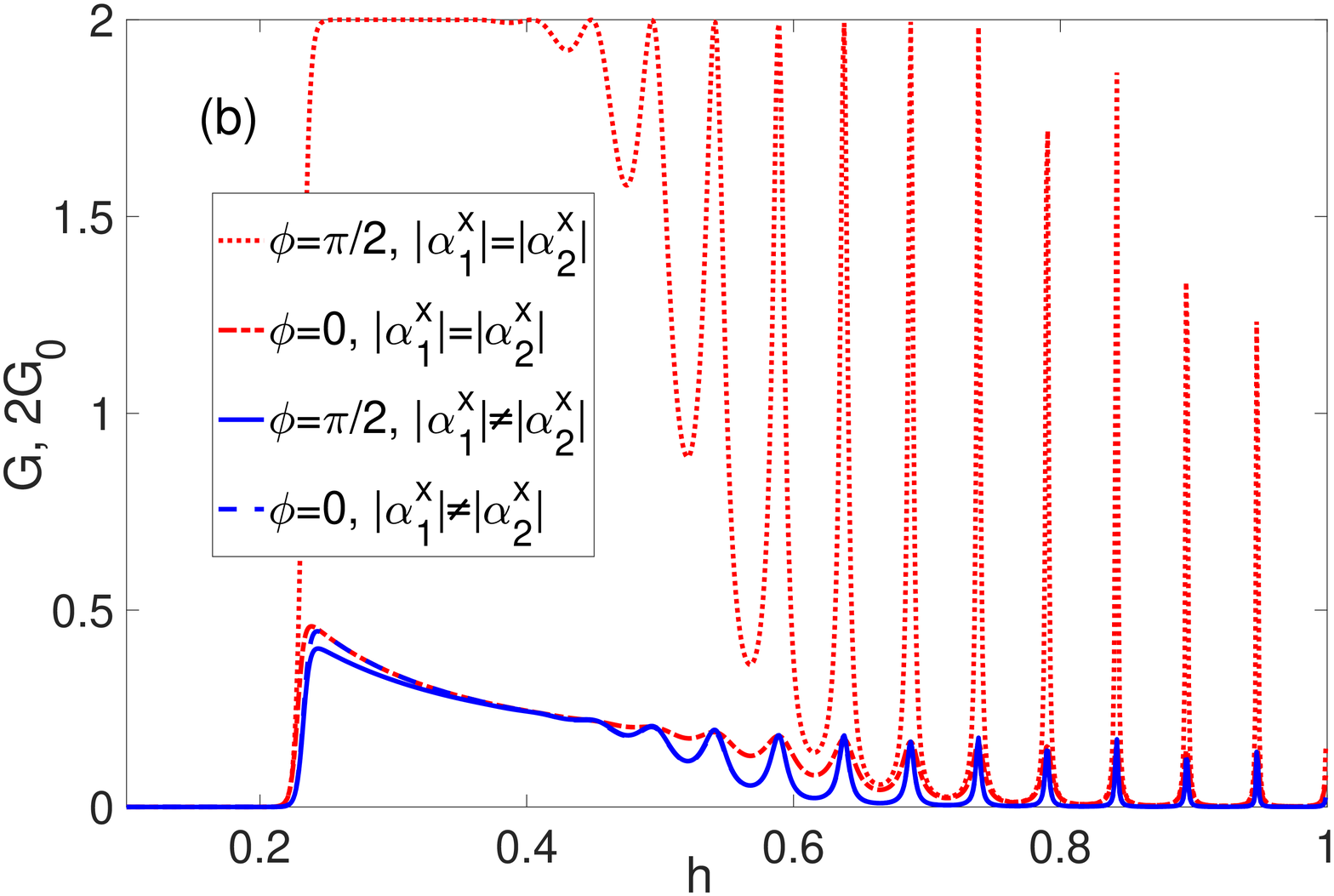}
	\caption{\label{fig2} Zeeman-field dependence of conductance for the zero (dashed curves) and nonzero (solid curves) Aharonov-Bohm phase if (a) $\alpha^{x}=0$ or (b) $\alpha^{x}\neq0$. The lowest excitation energy of the device as a function of the Zeeman energy is plotted in Fig. (a) by dotted curve (see right $y$ axis).}
\end{figure}

To extract the effect related to the presence of SC segment in the device on low-energy interference transport we start with the limiting case when $\alpha^{x}=0$. The corresponding Zeeman-energy dependence of conductance is shown in Fig. \ref{fig2}a. The behavior of $G\left(h\right)$ for the zero (dashed curve) and nonzero (solid curve) magnetic flux is qualitatively similar. At the low fields, $h<\Delta$, the SC wire is in the trivial phase and the energy of the device lowest state, $E_{1}$, is nonzero (see dotted curve) leading to $G=0$. In opposite, if $h>\Delta$ the phase is nontrivial. At the intermediate magnetic fields, close to the critical one, $E_{1}\approx0$ resulting in the monotonous dependence of $G\left(h\right)$. At $\phi=\pi/2$ the conductance reaches $4G_{0}$ value implying that both Majorana modes are involved in the transport between the normal and superconducting contacts contributing $2G_{0}$ each,
\begin{eqnarray}
&&b_{1}=\sum_{\sigma;j=1}^{N+2N_{1}}w_{j\sigma}\gamma_{j\sigma A},~b_{2}=\sum_{\sigma;j=1}^{N+2N_{1}}z_{j\sigma}\gamma_{j\sigma B},\label{b12}\\
&&w_{j\sigma}=v_{j\sigma}^{*}+u_{j\sigma},~z_{j\sigma}=i\left(v_{j\sigma}^{*}-u_{j\sigma}\right),\label{wz}
\end{eqnarray}
where $u_{j\sigma},~v_{j\sigma}$ - the spin-dependent Bogoliubov coefficients of the device lowest excitation; $\gamma_{j\sigma A\left(B\right)}=\gamma_{j\sigma A\left(B\right)}^{+}$ - the Majorana operators of type A(B). At the high fields the hybridization of Majorana wave functions becomes nonzero giving rise to the oscillations of $E_{1}\left(h\right)$ and $G\left(h\right)$. 

At zero AB phase the conductance is significantly suppressed. 
If the spin-orbit coupling in the arms is taken into account and $|\alpha_{1}^{x}|=|\alpha_{2}^{x}|$ that there is a slight quantitative difference from the $\alpha^{x}=0$ case (e.g., compare the solid curve in Fig. \ref{fig2}a with the dotted curve in Fig. \ref{fig2}b). When $|\alpha_{1}^{x}|\neq|\alpha_{2}^{x}|$ the divergence can be significant, which is especially prominent at $\phi=\pi/2$ (see solid curve in Fig. \ref{fig2}b). 

\begin{figure}[tb]
	\includegraphics[width=0.5\textwidth]{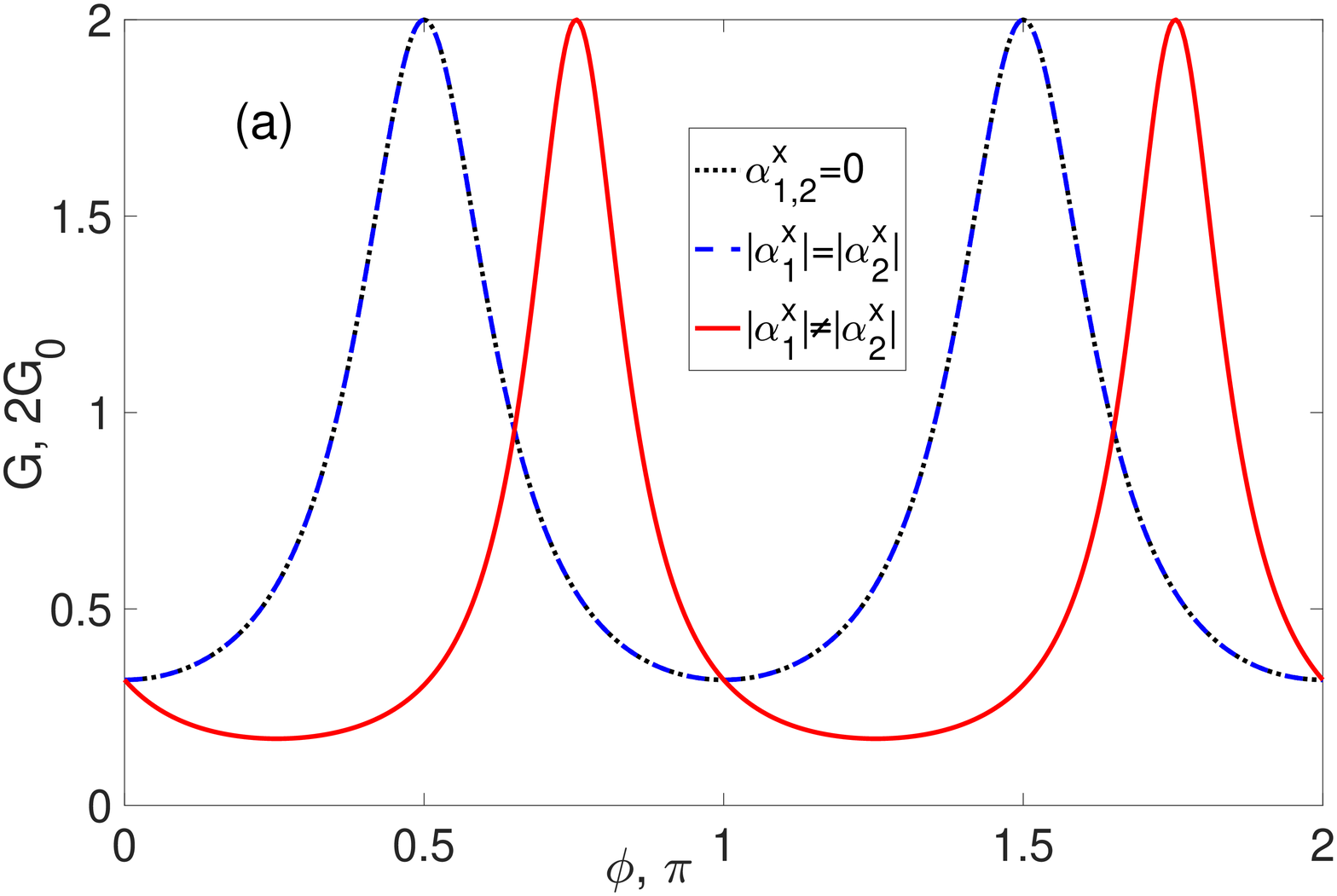}
	\includegraphics[width=0.5\textwidth]{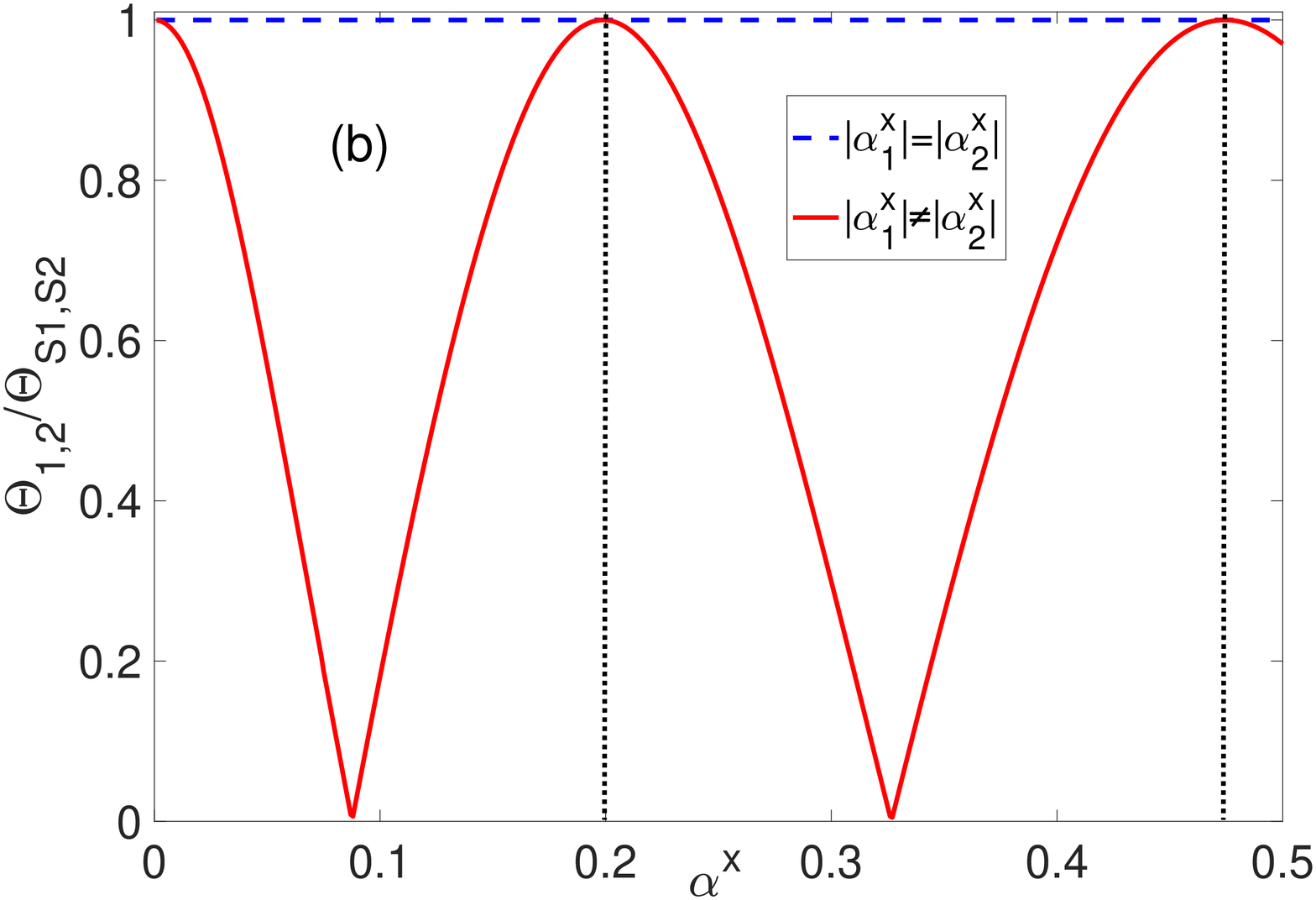}
	\caption{\label{fig3} (a) Aharonov-Bohm oscillations of conductance for nonoverlapping Majorana modes. (b) Ratio of angles between the spins at the edges of the device, $\Theta_{1,2}$, and at the opposite ends of the SC wire, $\Theta_{S1,S2}$, for the Majorana bound state. The  vertical dotted lines indicate the values of $\alpha^{x}$ for which the maxima of $G\left(\phi\right)$ occur at $\phi=\pi\left(n+1/2\right)$, $n\in \mathbb{Z}$. Parameters: $h=0.31$.}
\end{figure}

The AB oscillations of conductance are plotted in Fig.\ref{fig3}a when the device is in the nontrivial phase with the NOMMs. Both at $\alpha^{x}=0$ and at $\alpha^{x}\neq0$, $|\alpha_{1}^{x}|=|\alpha_{2}^{x}|$ the peaks appear when $\phi=\pi\left(n+1/2\right)$, $n\in \mathbb{Z}$ (see dotted and dashed curves). In turn, the different spin-orbit interaction in the arms leads to their shift as shown by solid curve. 

In general, spin-orbit coupling in the system results in electron spin precession around the Rashba vector. Then, an additional phase carriers acquire in such media propagating from $j$th to $j'$th site \cite{bercioux-05} can be related to the change of the angle between the spins of the device lowest excitation, $\psi_{1}$, at these sites, $\Theta\left( \mathbf{S}_{1}\left(j\right),\mathbf{S}_{1}\left(j'\right)\right)$, where	
\begin{equation}\label{Sxyze}
\mathbf{S}_{1}\left(j\right)=\psi_{1}^{+}\left(j\right)\bm{\sigma}\frac{\tau_{0}+\tau^{z}}{2}\psi_{1}\left(j\right).
\end{equation}	
Here $\bm{\sigma}=\left(\sigma^{x},\sigma^{y},\sigma^{z}\right),~\tau^{z}$ - the Pauli matrices acting in the spin and particle-hole space, respectively.

In the presence of mirror symmetry ($\alpha^{x}=0$ or $\alpha^{x}\neq0$, $|\alpha_{1}^{x}|=|\alpha_{2}^{x}|$) the angles between the spins at the opposite ends of the SC section, $\Theta\left( \mathbf{S}_{1}\left(j=N_{1}+1\right),\mathbf{S}_{1}\left(j=N_{1}+N\right) \right)\equiv\Theta_{S1,S2}$, and at the edges of the device, $\Theta\left( \mathbf{S}_{1}\left(j=1\right),\mathbf{S}_{1}\left(j=2N_{1}+N\right) \right)\equiv\Theta_{1,2}$, are the same. The latter is clearly seen in Fig. \ref{fig3}b where a ratio $\Theta_{1,2}/\Theta_{S1,S2}$ as a function of $\alpha^{x}$ is displayed (see dashed line). On the contrary, if $|\alpha_{1}^{x}|\neq|\alpha_{2}^{x}|$ the angles match only for the certain values of $\alpha^{x}$  (see vertical dotted lines in Fig. \ref{fig3}b). Outside these points of the parametric space the conductance resonances appear at $\phi\neq\pi(n+1/2)$ (see solid curve in Fig. \ref{fig3}a).
	
%The latter can be explained, for example, analyzing the $\alpha^{x}$-dependence of the spin of the device lowest excitation $\psi_{1}\left(j\right)$ at the top ($j=1$) and bottom ($j=N+2N_{1}$) edges. The electron component is defined as 
%\begin{equation}\label{Sxyze}
%\mathbf{S}_{1}\left(j\right)=\psi_{1}^{+}\left(j\right)\bm{\sigma}\frac{\tau_{0}+\tau^{z}}{2}\psi_{1}\left(j\right),
%\end{equation}
%where $\bm{\sigma}=\left(\sigma^{x},\sigma^{y},\sigma^{z}\right),~\tau^{z}$ - the Pauli matrices acting in the spin and particle-hole space, respectively. If $\alpha^{x}=0$ the spin has nonzero x- and z-projections only and $|S_{1}^{x,z}\left(1\right)|=|S_{1}^{x,z}\left(N+2N_{1}\right)|$ \cite{sticlet-12}. When $\alpha^{x}\neq0$ it goes out of the $xz$ plane and begins to rotate in the arm around the corresponding Rashba-field vector (these vectors in the different normal wires are oriented in the opposite directions) with the frequency $\sim\alpha^{x}/t$. In particular, it results in the oscillating  $\alpha^{x}$-dependence of $S_{1}^{z}\left(1\right)$ and $S_{1}^{z}\left(N+2N_{1}\right)$ depicted in Fig. \ref{fig3}b. The conductance shift disappears at $\alpha^{x}$ values such that $|S_{1}^{z}\left(1\right)|=|S_{1}^{z}\left(N+2N_{1}\right)|$ (see vertical dotted lines in Fig. \ref{fig3}b).   

\begin{figure*}[tb]
	\includegraphics[width=0.5\textwidth]{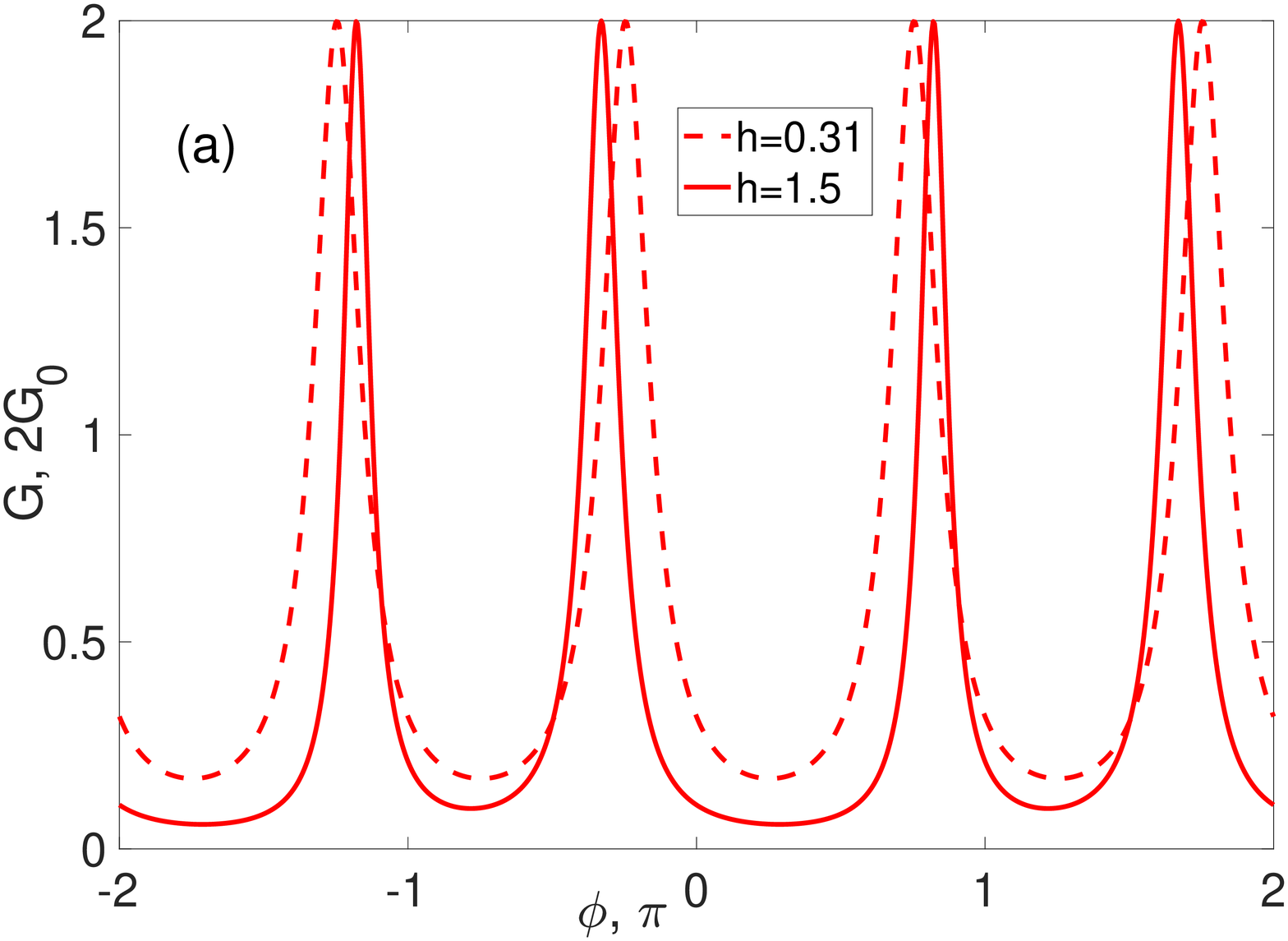}
	\includegraphics[width=0.5\textwidth]{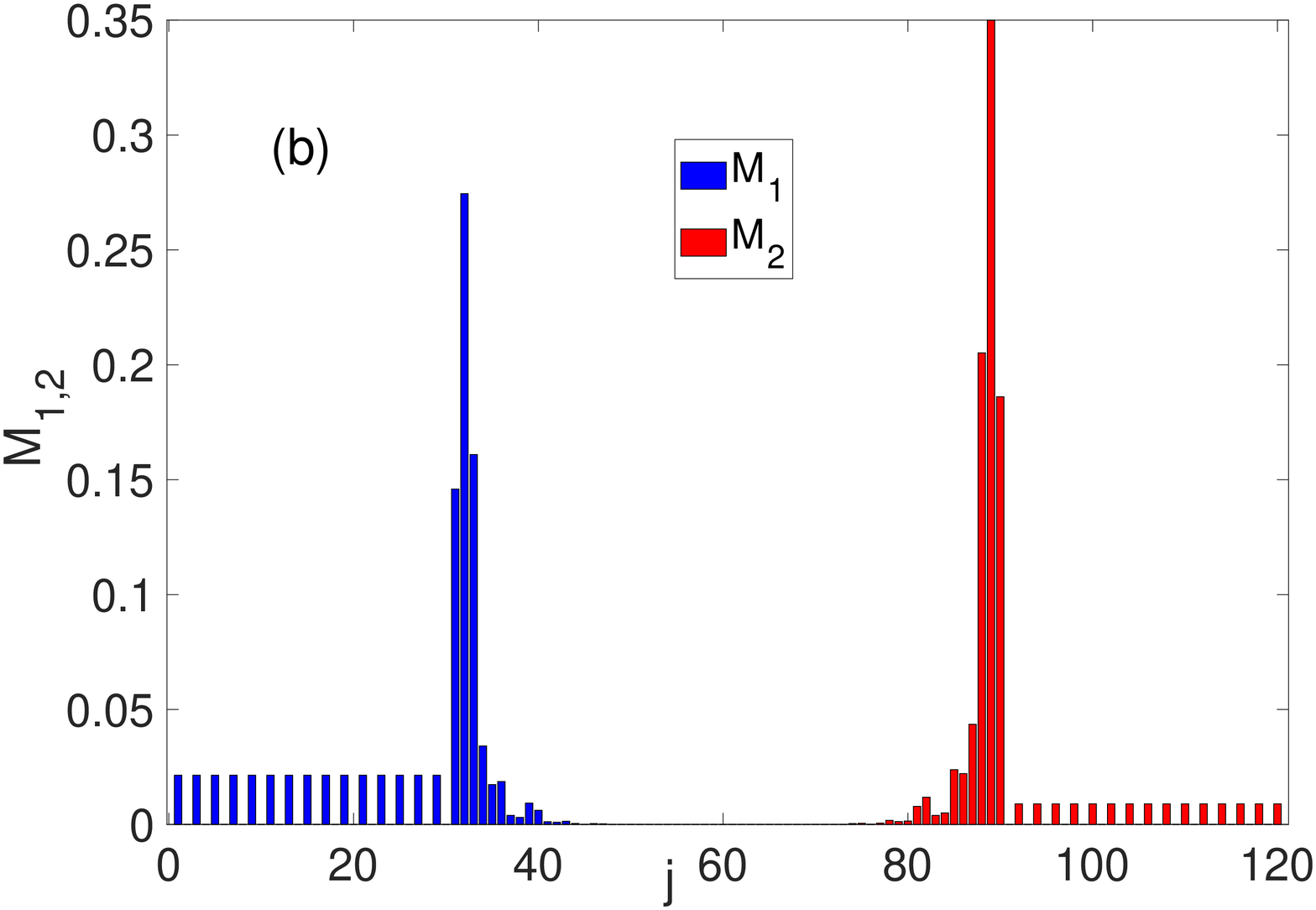}
	\includegraphics[width=0.5\textwidth]{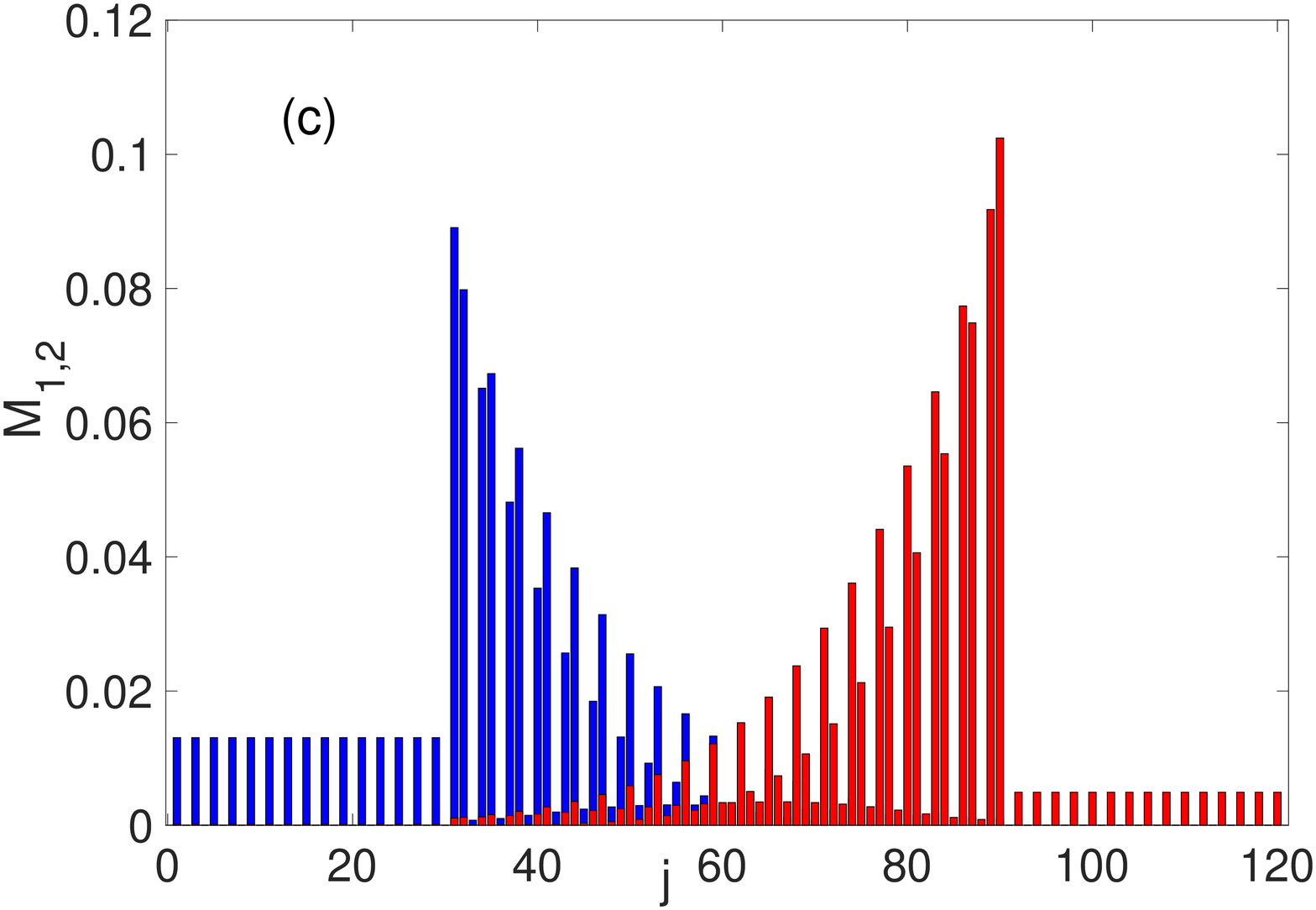}
	\includegraphics[width=0.52\textwidth]{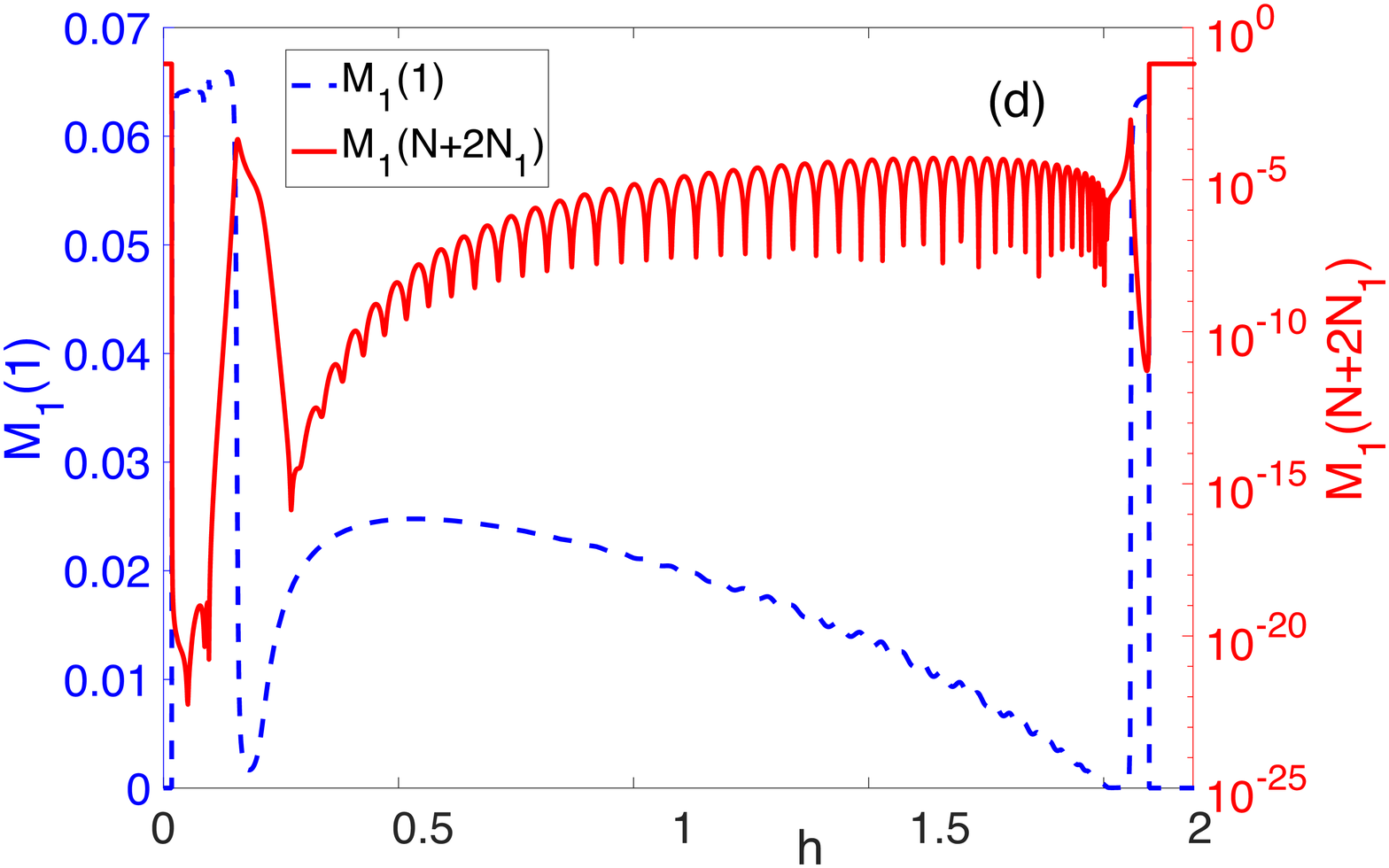}
	\caption{\label{fig4} (a) Effect of Majorana modes overlap on the Aharonov-Bohm oscillations of conductance. Spatial distribution of the Majorana modes at the intermediate, $h=0.31$ (b), and high, $h=1.5$ (c), magnetic fields. (d) Zeeman-energy dependence of  $b_{1}$ probability density at the top and bottom edge of the structure. Parameters: $|\alpha_{1}^{x}|\neq|\alpha_{2}^{x}|$.}
\end{figure*}

One of the main results of this study is displayed in Fig. \ref{fig4}a. As the Zeeman energy increases the positions of the maxima in the AB effect change and the oscillation period doubles. This process is accompanied by the increasing overlap of the Majorana wave functions. The corresponding probability densities, $M_{1}\left(j\right)=\sum_{\sigma}|w_{j\sigma}|^2$ and $M_{2}\left(j\right)=\sum_{\sigma}|z_{j\sigma}|^2$, are shown in Fig. \ref{fig4}b,c for $|\alpha_{1}^{x}|\neq|\alpha_{2}^{x}|$ at intermediate and high magnetic fields, respectively. To observe such a hybridization in transport the modes must have nonzero probability density at both ends of the device. In order for the leakage of Majorana modes \cite{vernek-14} into the arms to be possible in our system, the normal wires have to be quantum wells, $\varepsilon_{1,2}\leq\varepsilon$. Then, the desired situation of $M_{1}\left(N+2N_{1}\right),~M_{2}\left(1\right)\neq0$ realizes at the high magnetic fields as it can be seen from Fig.\ref{fig4}d where the $h$-dependencies of $M_{1}\left(1\right)$ and $M_{1}\left(N+2N_{1}\right)$ are plotted (the $b_{2}$ mode behaves identically). In the trivial phase ($h<0.2$ or $h>2$) the device lowest-energy excitation is localized in one of the arms since $|\alpha_{1}^{x}|\neq|\alpha_{2}^{x}|$ leading to either $M_{1}\left(1\right)=0$ or $M_{1}\left(N+2N_{1}\right)=0$. Immediately before the topological phase transition the $M_{1}\left(j\right)$ is mostly concentrated in the SC section causing its sudden drop at both edges of the device. In the nontrivial phase at the intermediate fields ($0.2<h\lesssim 0.5$) the $M_{1}\left(j\right)$ is localized predominantly at the boundary of the SC wire and arm $1$ (see Fig.\ref{fig4}b). Then, at the high fields, $0.5 \lesssim h<2$, the process of $b_{1}$ delocalization results in $M_{1}\left(N+2N_{1}\right)\neq0$ at the expense of the $M_{1}\left(1\right)$ reduction albeit $M_{1}\left(N+2N_{1}\right)\ll M_{1}\left(1\right)$. In turn, it leads to the narrowing of conductance resonant peaks (e.g. see Figs. \ref{fig2}b and \ref{fig4}a).

When the in-plane magnetic field increases the probability density in the central part of the SC wire becomes nonzero at the expense of decrease of the probability density at its ends and the arms.

Note that the described picture seems feasible in the modern experiments since to reach ballistic transport regime the short wires are used ($\sim 0.1 - 1~\mu$m). Taking additionally into account the available proximity-induced SC gaps ($\sim 0.1 - 1$ meV) one can conclude that the Majorana modes are able to extend over the whole structure \cite{yu-21,hess-21}. In the next section considering a spinless inference system we will analytically show that the direct interaction of the mode $b_{1\left(2\right)}$ with the arm $2\left(1\right)$ induces the AB peaks shift.

The proposed setup can be effective in distinguishing between the nontopological quasi-Majorana state and nonlocal topological Majorana state with spatially separated Majorana modes. The former emerge in the trivial phase due to the presence of smooth electrostatic and/or SC pairing potentials around the N/S interfaces \cite{fleckenstein-18,penaranda-18,vuik-19}. Despite the fact that in this case the overlap of the Majorana modes can vary significantly and be even comparable with the one in the nontrivial phase, they are still localized in such an inhomogeneous region. As a result, although the linear-response conductance can reach $4G_{0}$ for this type of states \cite{vuik-19} it is difficult to expect the observation of AB oscillations. On the other hand, the OMMs observed at the high magnetic fields can be treated as an ordinary fermionic subgap state featuring a bulk-like spatial distribution. As it is shown above the corresponding AB effect differs from the one for the nonlocal topological Majorana bound state, i.e. in the case of NOMMs.

%\begin{figure}[tb]
%	\includegraphics[width=0.45\textwidth]{fig1.eps}
%	\caption{\label{1} The Kitaev chain interacting with the normal contact via the single-level quantum dots.}
%\end{figure}

\section{\label{sec3}Analytical results for spinless model}

\subsection{\label{sec3.1} Formula for conductance mediated by local Andreev reflection}

To describe analytically the found features of low-energy interference transport let us proceed to the spinless system where the SC segment is represented by the Kitaev chain \cite{kitaev-01} and the arms are single-level quantum dots (QDs). The device Hamiltonian is 
\begin{eqnarray} \label{HDKit}
&&\hat{H}_{D} =\sum\limits_{i=1,2}\xi_{i}d^+_{i}d_{i}-\sum\limits_{j=1}^{N-1}\left(ta_{j}^{+}a_{j+1}-\Delta a_{j}^{+}a_{j+1}^{+}+h.c.\right)\nonumber\\
&&-\mu\sum\limits_{j=1}^{N}\left(a^+_{j}a_{j}-\frac{1}{2}\right)-\left(\tau_{1}d_{1}^{+}a_{1}+\tau_{2}d_{2}^{+}a_{N}+h.c.\right),
\end{eqnarray}
 
For further derivation of the conductance formula in the spinless case it is useful to separate electron and hole degrees of freedom in the normal-metal reservoir. In particular, the frequency-dependent blocks of the retarded and lesser self-energies in (\ref{IL3}) can be represented as 
\begin{eqnarray} \label{Sigmaeh}
&&\hat{\Sigma}_{ji}^{r}= -\frac{i}{2}diag\left(\Gamma_{eji},\Gamma_{hji}\right)\equiv-\frac{i}{2}\left(\hat{\Gamma}_{eji}+\hat{\Gamma}_{hji}\right),~\\
&&\hat{\Sigma}_{ji}^{+-}=
i\cdot diag\left(\Gamma_{eji}n_{e},\Gamma_{hji}n_{h}\right)\equiv i\left(\hat{\Gamma}_{eji}n_{e}+\hat{\Gamma}_{hji}n_{h}\right),\nonumber
\end{eqnarray} 
where $\Gamma_{ejj}=\Gamma_{j}$, $\Gamma_{e12}=\Gamma_{12}e^{i\phi}$, $\Gamma_{e21}=\Gamma_{e12}^{*}$, $\Gamma_{hji}=\Gamma_{eji}^{*}$. The last relations imply $\hat{\Gamma}_{e,h}=\hat{\Gamma}_{e,h}^{+}$ that, in turn, results in
\begin{equation} \label{SigG}
Tr\Bigl[\hat{\tilde{\sigma}}{\rm{Re}}\Bigl\{\hat{\Sigma}^{+-}\hat{G}^{a} \Bigr\} \Bigr]=-Tr\Bigl[\hat{\tilde{\sigma}}\hat{\Sigma}^{+-}\hat{G}^{r}\hat{\Sigma}^{r}\hat{G}^{a} \Bigr].
\end{equation}
Using the Keldysh equation and formula (\ref{SigG}) the current can be represented by the Landauer-type formula, $I=2\frac{e}{h}\int d\omega T_{LAR}\left(n_{h}-n_{e}\right)$, where the local Andreev reflection (LAR) probability is written as 
%\begin{equation} \label{TLAR1}
%T_{LAR}=
%\Tr\Bigl[{\rm{Re}}\Bigl\{\sum\limits_{\alpha=e,h}\hat{\Gamma}_{\alpha}\hat{G}^{r}\hat{\Gamma}_{\bar{\alpha}}\hat{G}^{a}\Bigr\} \Bigr]\equiv T_{eh}+T_{he}.
%\end{equation}
\begin{equation} \label{TLAR1}
T_{LAR}=
Tr\Bigl[{\rm{Re}}\Bigl\{\hat{\Gamma}_{e}\hat{G}^{r}\hat{\Gamma}_{h}\hat{G}^{a}\Bigr\} \Bigr].
\end{equation}
The factor $2$ in the expression for current appears since both electron and hole degrees of freedom participate at transport. Next, we appeal for the particular structure of the $\hat{G}^{r,a}$ blocks. After some mathematical calculations the LAR transmission becomes
%\begin{eqnarray} \label{TLAR2}
%&&T_{eh}=
%\mid e^{i\phi}\Gamma_{1}f_{e11}+e^{-i\phi}\Gamma_{2}f_{e22}+\Gamma_{12}\left(f_{e12}+f_{e21}\right)\mid ^2,~~~~~~\\
%&&T_{he}=\mid e^{-i\phi}\Gamma_{1}f_{h11}+e^{i\phi}\Gamma_{2}f_{h22}+\Gamma_{12}\left(f_{h12}+f_{h21}\right)\mid^2,\nonumber
%\end{eqnarray}
\begin{equation} \label{TLAR2}
T_{LAR}=
\mid e^{i\phi}\Gamma_{1}f_{11}+e^{-i\phi}\Gamma_{2}f_{22}+\Gamma_{12}\left(f_{12}+f_{21}\right)\mid ^2,
\end{equation}
where $f_{ji}\equiv\left\langle \left\langle d_{j}^{+}|d_{i}^{+}\right\rangle \right\rangle$ are the Fourier transforms of anomalous retarded GFs. For the sake of brevity, hereinafter the index 'r' denoting retarded character of the response is omitted. In the following we are concentrated on a study of quantum transport in the linear-response regime at the low-temperature limit. Then, the conductance is 
\begin{equation} \label{GLAR}
G=dI/dV=2G_{0}T_{LAR}.
\end{equation}
Thus, the analysis of transport features reduces to the finding of $f_{ji}$. In order to solve this problem the EOM technique is utilized for the three cases  \cite{zubarev-60}. The corresponding details are in \ref{apxA}.

\subsection{\label{sec3.2} Nonoverlapping Majorana modes}

As a starting point we consider the Kitaev chain at the symmetric point, $t=\Delta$ and $\mu=0$, with the adjacent QDs (the symmetric Kitaev chain, SKC). In this case there are two NOMMs.
The substitution of found GFs (\ref{aGF1}) into (\ref{TLAR2}) gives
\begin{eqnarray} \label{TLAR3}
&&T_{NOMM}=4t^4\omega^2\left[\Gamma_{1}^2\tau_{1}^4D_{2}^2+\Gamma_{2}^2\tau_{2}^4D_{1}^2-\right.\\
&&\left.~~~~~~~~~~~~~~~~~~~~~~-2\Gamma_{1}\Gamma_{2}\tau_{1}^2\tau_{2}^2D_{1}D_{2}\cos 2\phi\right]/\mid Z\mid^2,\nonumber
\end{eqnarray}
%$T_{he}=T_{eh}$
where $D_{j}=C\left(\omega^2-\varepsilon_{j}^2-\tau_{j}^2\right)-\tau_{j}^2\left(\omega^2-\tau_{j}^2\right)$, $C=\omega^2-4t^2$, $j=1,2$.

\begin{figure}[tb]
	\begin{center}
	\includegraphics[width=0.5\textwidth]{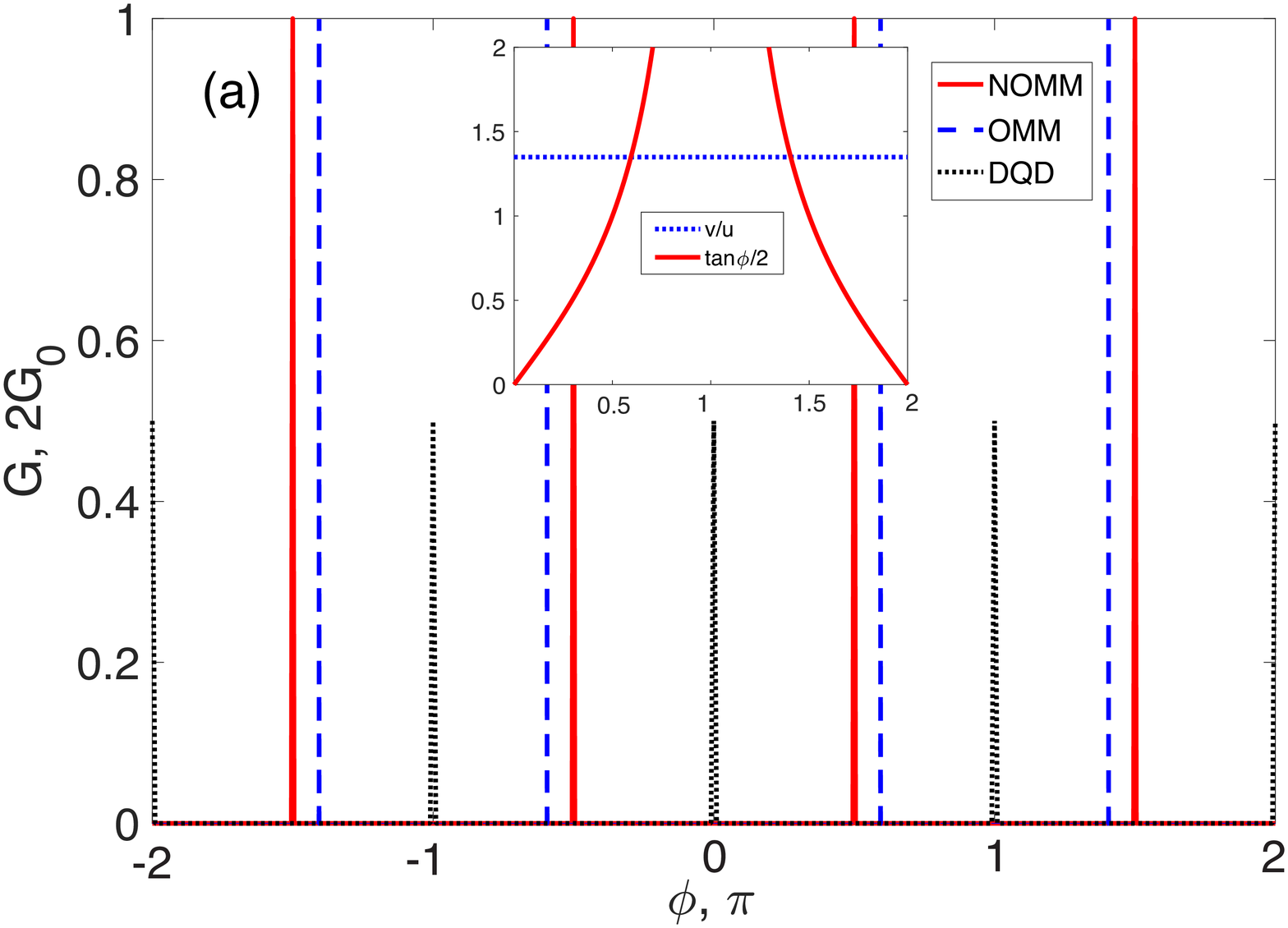}
	\includegraphics[width=0.525\textwidth]{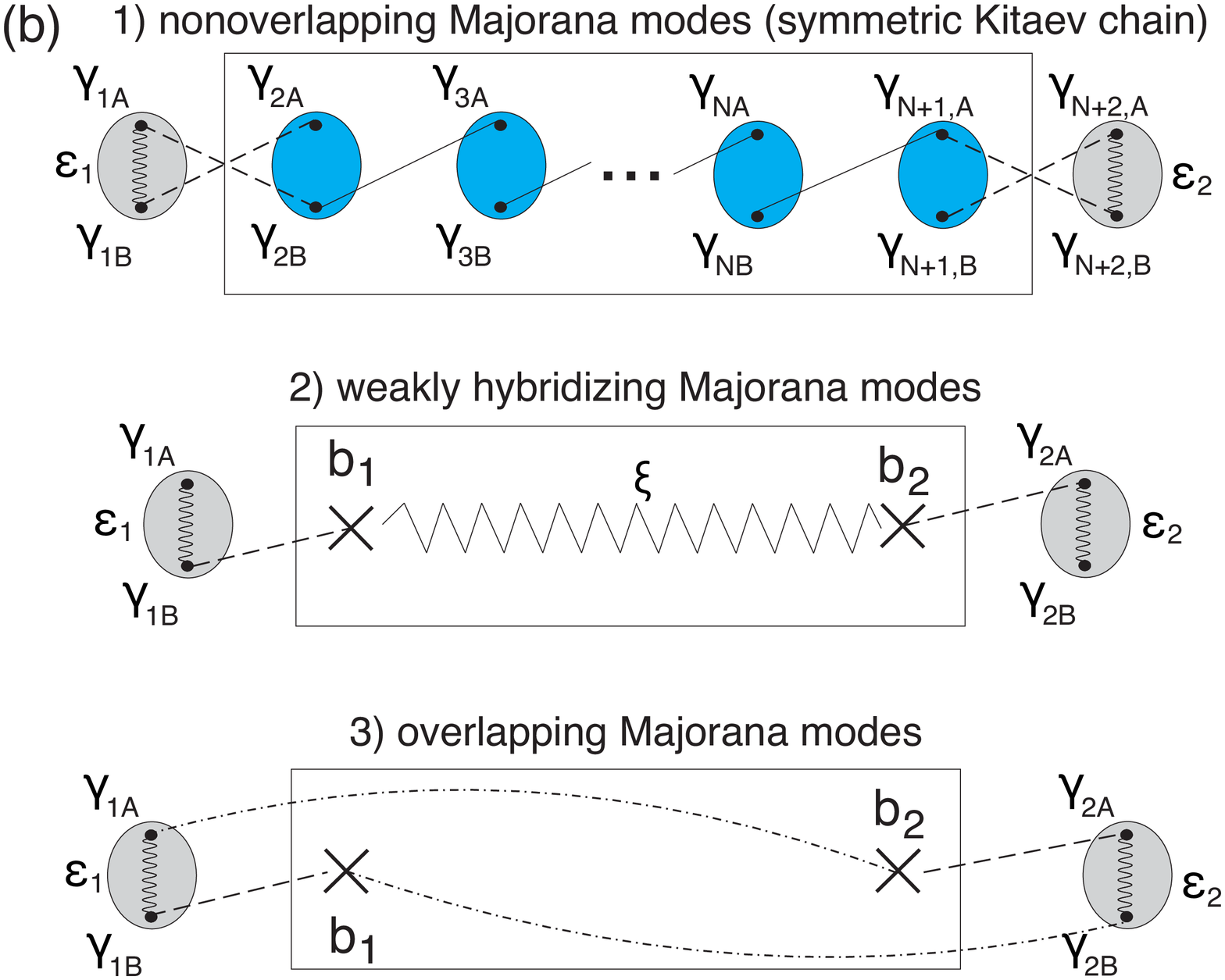}
	\caption{\label{2} (a) The linear-response conductance of the interference device as a function of the Aharonov-Bohm phase. The maxima emerge at $\phi=\pi \left(n+1/2\right)$ if the Kitaev chain is in the topological phase with the nonoverlapping Majorana modes (see solid curve). The peaks are shifted to the values $\phi=2\arctan v/u$ if the Majorana modes overlap (see dashed curve). Inset: the graphic solution of the equation $v/u=\tan\phi/2$. Parameters: $\Gamma_{1}=\Gamma_{2}=0.01$, $\tau_{1}=\tau_{2}=0.2$, $\varepsilon_{1}=\varepsilon_{2}=0$, $\Delta=t=1$, $\mu=0$, $N=20$. To obtain the zero-energy state with the overlapping Majorana modes $\mu\approx1.97$ and $\Delta=0.1$ are set. (b) Three types of the device in the Majorana representation.}
	\end{center}
\end{figure}

It is essential to notice that if $\phi=\pi \left(n+1/2\right)$, $n\in\mathbb{Z}$, the GFs have a first-order pole at $\omega=0$, i.e. $Z=\omega \tilde{Z}$. Then, 
\begin{equation} \label{TLAR4}
\tilde{T}_{NOMM}=4t^4\left[\Gamma_{1}\tau_{1}^2D_{2}+\Gamma_{2}\tau_{2}^2D_{1}\right]^2/\mid \tilde{Z}\mid^2,\nonumber
\end{equation}
where
\begin{eqnarray} \label{Z0}
&&\tilde{Z}=\omega\left\{D_{1}D_{2}-\frac{C}{4}\left(\Gamma_{1}^2D_{2}+\Gamma_{2}^2D_{1}+2\Gamma_{1}\Gamma_{2}D_{3}\right)\right\}+\nonumber\\
&&+i\Biggl\{2t^2\left(\Gamma_{1}\tau_{1}^2D_{2}+\Gamma_{2}\tau_{2}^2D_{1}\right)+\Biggr.\\
&&\Biggl.~~~~~~~~~~+\omega^2\left[\Gamma_{1}\left(C-\tau_{1}^2\right)D_{2}+\Gamma_{2}\left(C-\tau_{2}^2\right)D_{1}\right]\Biggr\},\nonumber
\end{eqnarray}
where $D_{3}=C\left(\omega^2-\varepsilon_{1}\varepsilon_{2}-\tau_{1}^2-\tau_{2}^2\right)+\tau_{1}^2\tau_{2}^2-2t^2\left(\tau_{1}^2+\tau_{2}^2\right)$. In other words, $\omega=0$ is an energy of the bound state in continuum. Taking into account the results (\ref{TLAR3}) and (\ref{TLAR4}) the conductance in the NOMM case is 
\begin{eqnarray} \label{G1}
G_{NOMM}=
\left\{\begin{array}{cc}
2G_{0},  & \phi=\pi \left(n+1/2\right); \\
0, & \phi\neq\pi \left(n+1/2\right).
\end{array}\right.~
\end{eqnarray}
The revealed behavior is universal as it is independent on the ratio between $\varepsilon_{j},~t_{j},~\tau_{j}$ ($j=1,2$). The obtained results are confirmed by numerics. In Fig. \ref{2}a the AB oscillations for the SKC case are displayed by the solid curve. 

The found features of conductance coincide with the ones for the situation when there is no QDs between the normal contact and Kitaev chain. The corresponding anomalous GF's are provided in \ref{apxA}2. Using them the LAR-mediated transmission can be expressed as
\begin{equation} \label{TSKC}
T_{0}=\frac{t^{4}\omega^{2}\left(\omega^{2}-t^{2}\right)^{2}}{4\mid Z\mid^2}\left[\Delta\Gamma^2\cos^{2}\phi+\Gamma^{2}\sin^{2}\phi\right],
\end{equation}
%$T_{he}=T_{eh}$
where $\Delta\Gamma=\Gamma_{1}-\Gamma_{2}$, $\Gamma=\Gamma_{1}+\Gamma_{2}$. Taking into account the phase dependence of the denominator, $Z$, presented in the formula (\ref{denGFnoQD}) the linear-response conductance is described by the expression (\ref{G1}) and is not affected by the $\Gamma_{1}/\Gamma_{2}$ ratio.

It is worth to note that the observed sharp AB oscillations resemble the corresponding conductance behavior of two noninteracting QDs  (double quantum dot, DQD) between normal contacts in the limiting case $\varepsilon_{1}=\varepsilon_{2}=0$ (see black dotted curve in Fig. \ref{2}a) \cite{kubala-02}. Thus, in the separate SKC (i.e. without the side-attached dots) two decoupled Majorana fermions, $\gamma_{1A}$ and $\gamma_{NB}$, can be treated as such a double-dot structure connected to the electron and hole reservoirs in parallel. However, it is essential that in the case of Majoranas the $G$ maxima are fundamentally shifted by the $\pi/2$ in comparison with the mentioned DQD situation \cite{kubala-02} (see blue dashed and black dotted curves in Fig. \ref{2}a). This effect is explained by the fact that the same phase carriers acquire tunneling from the contact to the $B$-type Majorana mode in the bottom arm. Note that the similar effect was demonstrated in the nonlocal transport through the QD with the side-coupled  SKC having the $\Pi$-shape \cite{liu-11,ramos-andrade-18}.

In general, the suggested analogy with two noninteracting QDs is not complete since in our case the other eigenstates having nonzero energies are also able to contribute to the transport. For example, in the simplest situation of SKC directly coupled with the contact two Majorana double-dot molecules (consisting of $\gamma_{1B},~\gamma_{2A}$ and $\gamma_{N-1,B},~\gamma_{NA}$, respectively) also interact with it. Then, both arms of the device include two chains characterized by opposite parity of Majorana-fermion number. Taking it into account in \ref{apxB} we numerically analyzed the transport properties of the analogous AB interferometer containing ordinary zero-energy QDs. It is shown the result (\ref{G1}) can be reproduced for any odd number of QDs in the separate arm (see \ref{apxB} for details). 

Hence, it becomes clear why there is no effect of nonzero $\varepsilon_{1},~\varepsilon_{2}$ on the conductance features, that can be perceived as another manifestation of the Majorana-mode leakage \cite{vernek-14}. The explicit reason of such a behavior is uncovered if the QD-SKC-QD device is considered in the Majorana-operator representation. As it is shown in the top row of Fig. \ref{2}b for any $\varepsilon_{1,2}$ there are two uncoupled chains characterized by an odd-parity number of the Majorana fermions connected in series. That, in turn, implies the presence of the two zero-energy eigenstates interacting with the normal contact. Obviously, the oscillations (\ref{G1}) have to keep for the arbitrary number of zero-energy QDs attached to the SKC edges.  

\subsection{\label{sec3.3} Weakly hybridizing Majorana modes}

Now we focus on the SC wire in the nontrivial phase modeled by the effective Hamiltonian $\hat{H}_{eff} =i\xi b_{1}b_{2}/2$, where $\xi\sim e^{-N}$ is a measure of hybridization of the Majorana modes $b_{1}=\sum\limits_{j=1}^{N}w_{j}\gamma_{jA}$ and $b_{2}=\sum\limits_{j=1}^{N}z_{j}\gamma_{jB}$ ($b_{1}=\gamma_{1A}$, $b_{2}=\gamma_{NB}$ if $N\to\infty$) \cite{kitaev-01}. As this interaction is supposed to be weak there is no immediate coupling between $b_{1}$ ($b_{2}$) and the right (left) QD  (see the middle row of Fig. \ref{2}b). Then, based on the necessary GFs which are brought in \ref{apxA}3, the conductance in the symmetric transport regime  (i.e. $\Gamma_{1}=\Gamma_{2}=\Gamma/2$, $\tau_{1}=\tau_{2}=\tau$, $\varepsilon_{1}=\varepsilon_{2}=\varepsilon$) has form
\begin{equation} \label{Teff}
	G_{eff}=2G_{0}\cdot\Gamma^{2}\tau^{4}\omega^2\frac{\left(\omega^2-\varepsilon^2-\tau^2\right)^2\sin^2\phi}{4\mid Z\mid^2}.
\end{equation}
%$T_{he}=T_{eh}$
From expression (\ref{Zeff}) one can easily conclude that for the NOMMs ($\xi=0$) the conductance at $\omega=0$ is equal to $2G_{0}$ if $\phi=\pi \left(n+1/2\right)$, $n\in\mathbb{Z}$. Otherwise, it is zero, that is in complete agreement with the above-obtained result (\ref{G1}). However, the mentioned features also take place if $\varepsilon=0$ even though $\xi\neq0$. The latter can be easily explained if one again refers to the system in the Majorana representation displayed in the middle row of Fig. \ref{2}b. Indeed, when on-dot energies $\varepsilon_{1,2}$ are zero two unpaired Majorana fermions, $\gamma_{1A}$ and $\gamma_{2B}$, occur giving rise to one zero-energy state in each arm. In general situation of $\varepsilon,~\xi\neq0$ the linear-response conductance is zero. Mathematically, it means that the first-order pole of the GFs arises at $\omega=0$ only if $\cos\phi=2\varepsilon\xi\left(\pm 2i\varepsilon/\Gamma - 1 \right)/\tau^2$.

\subsection{\label{sec3.4} Overlapping Majorana modes}

Finally, far from the symmetric point of the Kitaev model, the Majorana zero modes become to hybridize significantly. In particular, it infers that the mode, $b_1$ or $b_2$, initially localized at one of the chain's edges shows the nonzero probability density at the opposite edge. It results in the direct coupling to both QDs with the different amplitudes proportional to $\tau_{1,2}\left(u \pm v\right)$ or $\tau_{1,2}\left(u \mp v\right)$ (see the dashed and dash-dotted curves in the bottom row of Fig. \ref{2}b, respectively) \cite{prada-17,clarke-17}. If $t/\Delta\gg1$ there are $N$ values of the chemical potential (\ref{mu0eq}) at which the zero-energy Bogoliubov excitation with the OMMs emerges. Solving the EOMs (\ref{Sys_GFABS}) presented in \ref{apxA}4 one can obtain the following transmission in the symmetric transport regime:
\begin{equation} \label{TABS}
T_{OMM}=\frac{4\Gamma^{2}\tau_{e}^{2}\tau_{h}^{2}\omega^{2}\left(\omega^2-\varepsilon^2-2\tau_{p}\right)^{2}\sin^{2}\phi}{\mid Z\mid^2}.
\end{equation}
%$T_{he}=T_{eh}$
Taking into account the $\omega$-dependence of the GFs' denominator (\ref{ZABS}) the linear-response conductance is 
\begin{eqnarray} \label{GABS}
	G_{OMM}=
	\left\{\begin{array}{cc}
		2G_{0},  & \varepsilon=0~\mbox{and}~v/u= \tan\phi/2; \\
		0, & \varepsilon\neq0~\mbox{or}~v/u\neq \tan\phi/2.
	\end{array}\right.~
\end{eqnarray}
The AB oscillations are depicted in Fig. \ref{2}a by the dashed curve. According to the expression (\ref{GABS}) the maxima appear at the phase values $\phi=2\arctan v/u$ (see the graphic solution of the equation $v/u= \tan\phi/2$ in the inset). Note that the positions of conductance peaks tend to $\phi=\pi \left(n+1/2\right)$ if the hybridization of the Majorana modes vanishes meaning the condition $u=v$. Hence, in the OMM situation the AB oscillation period doubles (compare solid and dashed curves in Fig.\ref{2}a).

The difference between $\varepsilon\neq0$ and $\varepsilon=0$ cases can be readily understood looking at the bottom row of Fig. \ref{2}b. If $\varepsilon=0$ two noninteracting chains emerge each of which is a triple Majorana-dot molecule coupled in parallel with the electron and hole reservoirs. In other words, it again provides the doubly degenerate zero-energy eigenstate of the device. Such a state is absent when $\varepsilon\neq0$.   

\section{\label{sec4}Summary}

In this study we offered a new method to examine the overlap of Majorana modes in the hybrid structure containing the SC segment between the normal wires (arms). In particular, the AB oscillations of linear-response conductance are studied when the normal contact is tunnel coupled with both arms (the $\Pi$-shape geometry). As soon as the hybridization of Majorana modes becomes significant the AB resonances shift and the AB oscillation period doubles. The effect emerges since each of the overlapping modes interact with the contact via two arms that is supported by the numerical calculations for the initial spinful model and the analytical analysis for the spinless system including the Kitaev chain and QDs.

Additionally, it is shown that if the factor of spin-orbit interaction in the arms is excluded the conductance peaks related to the perfectly delocalized Majorana bound state occur at $\phi=\pi\left(n+1/2\right)$, $n\in \mathbb{Z}$. It differs from the resonant AB phase $\phi=\pi n$ of the usual interferometer including two uncoupled arms with the odd number of ordinary zero-energy QDs between normal contacts \cite{kubala-02}. The reason of this mismatch is that in the former, in fact, the device is coupled to the electron and hole reservoirs by two different-type Majorana operators. By definition, the tunneling phase into them differs by $\pi/2$. In the latter such a feature is absent since the interaction between the contacts and each arm is implemented via the Majorana operators of both types.

\ack
We acknowledge fruitful discussions with A. D. Fedoseev and M. S. Shustin. The reported study was funded Russian Foundation for Basic Research (project No 20-02-00015), Government of Krasnoyarsk Territory, Krasnoyarsk Regional Fund of Science to the research project "Study of edge states in one- and two-dimensional topological superconductors" (No. 20-42-243005).
% and the Council of the President of the Russian Federation for Support of Young Scientists and Leading Scientific Schools, project No. MK-1641.2020.2. 

\appendix

\section{\label{apxA} The Green's functions of the spinless interference device}

In this Appendix the GFs for different SC interference devices coupled to the single normal contact in the spinless regime are brought. To obtain them the EOM technique is applied. A general form of the EOMs for the normal and anomalous retarded frequency-dependent GFs, $\left\langle \left\langle d_{j}|d_{i}^{+}\right\rangle \right\rangle\equiv g_{ji}$ and  $\left\langle \left\langle d_{j}^{+}|d_{i}^{+}\right\rangle \right\rangle\equiv f_{ji}$, is
\begin{eqnarray} \label{eqG1}
&&\left(\omega+i\delta\right)\langle\langle d_{j} | d_{i}^{+} \rangle\rangle=\left\langle\left\{d_{j},~d_{i}^{+}\right\}\right\rangle+
\langle\langle \left[d_{j},~\hat{H}\right] | d_{i}^{+} \rangle\rangle,\nonumber\\
&&\left(\omega+i\delta\right)\langle\langle d_{j}^{+} | d_{i}^{+} \rangle\rangle=
\langle\langle \left[d_{j}^{+},~\hat{H}\right] | d_{i}^{+} \rangle\rangle.
\end{eqnarray}
The EOM system solution makes it possible to find transmission amplitude associated with LAR processes since
\begin{equation} \label{tLAR}
t_{LAR}=
e^{i\phi}\Gamma_{1}f_{11}+e^{-i\phi}\Gamma_{2}f_{22}+\Gamma_{12}\left(f_{12}+f_{21}\right).
\end{equation}
Thus, the following subsections present the diagonal and off-diagonal anomalous GFs.

\textit{1. The symmetric Kitaev chain}

In order to treat the transport problem analytically we first address to the Kitaev model at the symmetric point, $t=\Delta$ and $\mu=0$. Employing the Majorana-operator representation, $a_{j}=\frac{1}{2}\left(\gamma_{jA}+i\gamma_{jB}\right)$, where $\gamma_{jA,B}=\gamma_{jA,B}^{+}$, one can rewrite the device Hamiltonian (\ref{HDKit}) in the form
\begin{eqnarray} \label{HD2} 
	&&\hat{H}_{D} =\sum\limits_{i=1,2}\xi_{i}d^+_{i}d_{i}+it\sum\limits_{j=1}^{N-1}\gamma_{jB}\gamma_{j+1,A}-\nonumber\\
	&&~~~~~~~~-\frac{\tau_{1}}{2}\left[\left(d_{1}^{+}-d_{1}\right)\gamma_{1A}+i\left(d_{1}^{+}+d_{1}\right)\gamma_{1B}\right]-\\
	&&~~~~~~~~~~~~~~~-\frac{\tau_{2}}{2}\left[\left(d_{2}^{+}-d_{2}\right)\gamma_{NA}+i\left(d_{2}^{+}+d_{2}\right)\gamma_{NB}\right].\nonumber
\end{eqnarray}

As a result, the EOM system (\ref{eqG1}) for $i=1$ is $\hat{A}_{1}\hat{x}_{1}=\hat{B}_{1}$, where $\hat{x}_{1}=\left(g_{11},g_{21},f_{11},f_{21}\right)^{T}$. The coefficient matrix and column vector are
\begin{eqnarray}
\label{Sys_GF1}
&&\hat{A}_{1}=\left({\begin{array}{*{4}{c}}
	z_{1e}&z_{12}&2t^2\tau_{1}^2&0\\
	-z_{12}^{*}&z_{2e}&0&-2t^2\tau_{2}^2\\
	2t^2\tau_{1}^2&0&z_{1h}&-z_{12}^{*}\\
	0&-2t^2\tau_{2}^2&z_{12}&z_{2h}
	\end{array}} \right),\nonumber\\ 
&&\hat{B}_{1}=\left(\omega C,0,0,0\right)^{T},
\end{eqnarray}
where $z_{je\left(h\right)}=C\left(\omega C_{je\left(h\right)}-\tau_{j}^2\right)-2t^2\tau_{j}^2$, $z_{12}=i\omega C\Gamma_{12}e^{i\phi}/2$, $C=\omega^2-4t^2$, $C_{je\left(h\right)}=\omega\mp\varepsilon_{j}+i\Gamma_{j}/2$.

The EOM system for $i=2$ can be written in a similar manner. Their solution gives 
the sought anomalous GFs,
\begin{eqnarray}
\label{aGF1}
&&f_{11}=-\frac{t^2}{2}\cdot\frac{\omega C\tau_{2}^2 \Gamma_{12}^2e^{-2i\phi}+4\tau_{1}^2Z_{2}}{Z},\nonumber\\
&&f_{22}=\frac{t^2}{2}\cdot\frac{\omega C\tau_{1}^2 \Gamma_{12}^2e^{2i\phi}+4\tau_{2}^2 Z_{1}}{Z},\nonumber\\
&&f_{21}=i\Gamma_{12}t^2\cdot\frac{\tau_{1}^2 z_{2e}e^{i\phi}-\tau_{2}^2 z_{1h}e^{-i\phi}}{Z},\\
&&f_{12}=i\Gamma_{12}t^2\cdot\frac{\tau_{1}^2 z_{2h}e^{i\phi}-\tau_{2}^2 z_{1e}e^{-i\phi}}{Z},\nonumber
%&&f_{hjj}=f_{ejj}\left(\phi\to-\phi\right),~f_{hji}=f_{eij}\left(\phi\to-\phi\right),\nonumber
\end{eqnarray}
where
\begin{eqnarray}
\label{ZZ}
&&Z_{j}=\omega\left(CC_{je}C_{jh}+\tau_{j}^4\right) -\tau_{j}^2(C+2t^2)(C_{je}+C_{jh}),\nonumber\\
&&Z=Z_{1}Z_{2}+\frac{\Gamma_{12}^2}{4}\Bigl(z_{1e}z_{2e}+z_{1h}z_{2h}+\omega^2 C^2\frac{\Gamma_{12}^2}{4}-\Bigr.\\
&&\Bigl.~~~~~~~~~~~~~~~~~~~~~~~~~~~~~~~~~~~~~~~~~~~~~~-8t^4\tau_{1}^2 \tau_{2}^2 \cos 2\phi \Bigr).\nonumber
\end{eqnarray}

\textit{2. The symmetric Kitaev chain directly coupled to the normal contact}\\

If there are no dots between the contact and Kitaev chain, then in the symmetric point one can easily write the following coefficient matrix and column vector writing the EOMs for the Majorana GFs, $\hat{A}_{1A}\hat{x}_{1A}=\hat{B}_{1A}$, where $\hat{x}_{1A}=\left(m_{1A,1A},m_{1B,1A},m_{NA,1A},m_{NB,1A}\right)^{T}$ and $m_{iA,jB}=\left\langle \left\langle \gamma_{iA}|\gamma_{jB}\right\rangle \right\rangle$:
\begin{eqnarray}
	\label{Sys_GF_Maj}
	%\hat{A}_{1A}=\left({\begin{array}{*{4}{c}}
	%	\Omega_{1}&0&2i\Gamma_{12}\omega\cos\phi&-2i\Gamma_{12}\omega\sin\phi\\
	%	0&\omega+2i\Gamma_{1}&2i\Gamma_{12}\omega\sin\phi&2i\Gamma_{12}\omega\cos\phi\\
	%	2i\Gamma_{12}\omega\cos\phi&2i\Gamma_{12}\omega\sin\phi&\omega+2i\Gamma_{2}&0\\
	%	-2i\Gamma_{12}\omega\sin\phi&2i\Gamma_{12}\omega\cos\phi&0&\Omega_{2}
	%	\end{array}} \right),\nonumber\\ 
	&&\hat{A}_{1A}=\left({\begin{array}{*{2}{c}}
			\hat{D}_{1}&\hat{T}_{12}\\
			\hat{T}^{T}_{12}&\hat{D}_{2}
	\end{array}} \right),
	\hat{T}_{12}=2i\Gamma_{12}\omega\left({\begin{array}{*{2}{c}}
			\cos\phi&-\sin\phi\\
			\sin\phi&\cos\phi
	\end{array}} \right),\nonumber\\ &&\hat{D}_{1}=diag\left(\Omega_{1},\omega_{1}\right),~\hat{D}_{2}=diag\left(\omega_{2},\Omega_{2}\right),\nonumber\\ 
	&&\hat{B}_{1A}=\left(2\omega,0,0,0\right)^{T},
\end{eqnarray}
where $\Omega_{j}=\omega\omega_{j}-t^2$, $\omega_{j}=\omega+2i\Gamma_{j}$. Note that the factor $2$ in $\hat{B}_{1A}$ is caused by the nonfermionic anticommutation relations, i.e. $\{\gamma_{iA\left(B\right)},\gamma_{jA\left(B\right)}\}=2\delta_{ij}$. The three systems for the rest of Majorana GFs, i.e. $m_{jA\left(B\right),1B}$, $m_{jA\left(B\right),2A}$ and $m_{jA\left(B\right),2B}$, have similar form. Finally, taking into account the relation between the Majorana and fermionic GF's,
\begin{equation} \label{FerMajGF}
f_{ij}=\frac{1}{4}\left[m_{iA,jA}-m_{iB,jB}- i\left(m_{iA,jB}+m_{iB,jA}\right)\right],
\end{equation}
the required anomalous GFs are
\begin{eqnarray}
\label{aGFnoQD}
&&f_{11}=-\frac{t^2}{2}\cdot \frac{\Omega_{2}\left(\omega+2i\Gamma_{2}\right)+4\Gamma_{12}^{2}\omega e^{2i\phi}}{Z},\nonumber\\
&&f_{22}=\frac{t^2}{2}\cdot \frac{\Omega_{1}\left(\omega+2i\Gamma_{1}\right)+4\Gamma_{12}^{2}\omega e^{-2i\phi}}{Z},\\
&&f_{12}=f_{21}=\Gamma_{12}t^2\cdot\frac{2\Delta\Gamma\omega\cos\phi+\left[\omega\left(\omega_{1}+\omega_{2}\right)-t^2\right]\sin\phi}{Z},\nonumber\\
&&Z=\omega\Omega\left[\omega\Omega-2i\Gamma t^2\right]-4\Gamma_{12}^{2}t^4\cos^{2}\phi,\label{denGFnoQD}
\end{eqnarray}
%\begin{eqnarray}
%	\label{MajGFs}
%	&&m_{1A,1A\left(NB,NB\right)}=\frac{2}{Z}\Bigl\{\Omega_{1}\Omega_{2}\omega_{2,1}+\Bigr.\\
%	&&\Bigl.~~~~~~~~~~~~~~~~~~~~~~~~~~~~~~~+4\Gamma_{12}^{2}\omega\left[\omega\omega_{2,1}-t^2\cos^{2}\phi\right]\Bigr\},\nonumber\\
%	&&m_{1B,1B\left(NA,NA\right)}=\frac{2\omega}{Z}\Bigl\{\Omega_{2,1}\omega_{1}\omega_{2}+\Bigr.\nonumber\\
%	&&\Bigl.~~~~~~~~~~~~~~~~~~~~~~~~~~~~~~~~~+4\Gamma_{12}^{2}\left[\omega\omega_{2,1}-t^2\sin^{2}\phi\right]\Bigr\},\nonumber\\
%	&&m_{1A,NB}=m_{NB,1A}=-\frac{4i\Gamma_{12}}{Z}\left(\Omega_{1}\Omega_{2}+4\Gamma_{12}^{2}\omega^{2}\right)\sin\phi,\nonumber\\
%	&&m_{1A,1B\left(NA,NB\right)}=m_{1B,1A\left(NB,NA\right)}=\frac{4\Gamma_{12}^{2}t^2}{Z}\omega\sin^{2}2\phi,\nonumber\\
%	&&m_{1B,NA}=m_{NA,1B}=\frac{4i\Gamma_{12}}{Z}\omega^{3}\left(\omega+2i\Gamma\right)\sin\phi,\nonumber\\
%	&&m_{1A,NA\left(1B,NB\right)}=-\frac{4i\Gamma_{12}}{Z}\omega\left(\Omega_{1,2}\omega_{2,1}+4\Gamma_{12}^{2}\omega\right)\cos\phi,\nonumber\\
%	&&m_{NA,1A\left(NB,1B\right)}=m_{1A,NA\left(1B,NB\right)},\nonumber\\ \label{denMajGFs}
%	&&Z=\omega\Omega\left[\omega\Omega-2i\Gamma t^2\right]-4\Gamma_{12}^{2}t^4\cos^{2}\phi,\label{ZMajGFs}
%\end{eqnarray}
where $\Gamma=\Gamma_{1}+\Gamma_{2}$, $\Delta\Gamma=\Gamma_{1}-\Gamma_{2}$, $\Omega=\omega^{2}+2i\Gamma\omega-t^{2}$.\\ 

\textit{3. The effective model of the Kitaev chain}\\

Let us turn to the effective model of the Kitaev chain in the nontrivial phase which describes two Majorana modes localized at the opposite ends on the chain and weakly interacting with each other \cite{kitaev-01}. In this case the device Hamiltonian can be written as
%\begin{eqnarray} \label{HDeff}
%&&\hat{H}_{D} =\sum\limits_{i=1,2}\xi_{i}d^+_{i}d_{i}+i\frac{\xi}{2}\gamma_{1}\gamma_{2}-\\
%&&~~~~~~~~~~~~~~~~~~~~~~-\frac{\tau_{1}}{2}\left(d_{1}^{+}-d_{1}\right)\gamma_{1}-\frac{\tau_{2}}{2}\left(d_{2}^{+}-d_{2}\right)\gamma_{2},\nonumber
%\end{eqnarray}
\begin{eqnarray} \label{HDeff}
	&&\hat{H}_{D} =\sum\limits_{i=1,2}\xi_{i}d^+_{i}d_{i}+i\frac{\xi}{2}b_{1}b_{2}-\\
	&&~~~~~~~~~~~~~~~~~~~~~~-\frac{\tau_{1}}{2}\left(d_{1}^{+}-d_{1}\right)b_{1}-i\frac{\tau_{2}}{2}\left(d_{2}^{+}+d_{2}\right)b_{2},\nonumber
\end{eqnarray}
where $\xi$ - a hybridization parameter. Hence, the EOM technique leads to 
%\begin{eqnarray}
%\label{Sys_GFeff}
%\hat{A}_{e1}=\left( {\begin{array}{*{10}{c}}
%		z_{1e}& Z_{12}&T_{1}&T_{12}\\
%		-Z_{12}^{*}+2T_{12}&z_{2e}&- T_{12}&T_{2}\\
%		T_{1}& T_{12}&z_{1h}&-Z_{12}^{*}\\
%		-T_{12}&T_{2}&Z_{12}+2T_{12}&z_{2h}
%\end{array}} \right)\nonumber\\ 
%\hat{b}_{e1}=\left(\omega^2-\xi^2,0,0,0\right)^{T},
%\end{eqnarray}
%where $T_{12}=i\xi\tau_{1}\tau_{2}/2$, $T_{j}=\omega\tau_{j}^2/2$, $Z_{12}=i\left(\omega^2-\xi^2\right)\Gamma_{12}e^{i\phi}/2-T_{12}$.
\begin{eqnarray}
	\label{Sys_GFeff}
	&&\hat{A}_{1}=\left( {\begin{array}{*{10}{c}}
			z_{1e}& Z_{12}&T_{1}&T_{12}\\
			-Z_{12}^{*}&z_{2e}&- T_{12}&T_{2}\\
			T_{1}& -T_{12}&z_{1h}&-Z_{12}^{*}-2T_{12}\\
			T_{12}&T_{2}&Z_{12}-2T_{12}&z_{2h}
	\end{array}} \right)\nonumber\\ 
	&&\hat{B}_{1}=\left(\omega^2-\xi^2,0,0,0\right)^{T},
\end{eqnarray}
where $z_{1e\left(h\right)}=\left(\omega^2-\xi^2\right)C_{1e\left(h\right)}-T_{1}$, $z_{2e\left(h\right)}=\left(\omega^2-\xi^2\right)C_{2e\left(h\right)}-T_{2}$, $T_{12}=-\xi\tau_{1}\tau_{2}/2$, $T_{j}=\omega\tau_{j}^2/2$, $Z_{12}=i\left(\omega^2-\xi^2\right)\Gamma_{12}e^{i\phi}/2+T_{12}$.

Since in the general case $\hat{A}_{1}$ has no zero elements the solution of EOM system $\hat{A}_{1}\hat{x}_{1}=\hat{B}_{1}$ is more cumbersome. We adduce it for the symmetric transport configuration when $\Gamma_{1}=\Gamma_{2}=\Gamma/2$, $\tau_{1}=\tau_{2}=\tau$, $\varepsilon_{1}=\varepsilon_{2}=\varepsilon$,
%\begin{eqnarray}
%\label{aGFeff}
%&&f_{e\left(h\right)11}=\frac{\tau^2}{4Z}\left[4\omega\left(\omega^2-\tau^2-\varepsilon^2\right)+\right.\nonumber\\ 
%&&\left. ~~~~~~~~~~~~+ i\Gamma\left(2\omega^2-\tau^2- 2\varepsilon\xi e^{\mp i\phi}\mp\frac{\Gamma}{2}\omega e^{\mp i\phi}\sin\phi\right) \right],\nonumber\\
%&&f_{e21\left(12\right)}=\mp\frac{\tau^2}{4Z}\left[4\xi\left(\omega^2-\varepsilon^2+i\Gamma\omega\right)-2i\Gamma\varepsilon\omega\cos\phi\mp\right.\nonumber\\
%&&\left. ~~~~~~~~~~~~~~~~~~~~~~~~~~~~\mp \Gamma\left(\omega\left(2\omega+i\frac{\Gamma}{2}\right)-\tau^2 \right) \sin\phi \right],\nonumber\\
%&&f_{e\left(h\right)22}=-f_{h\left(e\right)11},~f_{h21\left(12\right)}=-f_{e21\left(12\right)},
%\end{eqnarray}
\begin{eqnarray}
	\label{aGFeff}
	&&f_{11\left(22\right)}=\pm\frac{\tau^2}{4Z}\Biggl[4\omega\left(\omega^2-\tau^2-\varepsilon^2\right)+\Biggr. \\ 
	&&\Biggl. ~~~~~~~~~~~~+ i\Gamma\left(2\omega^2-\tau^2- 2\varepsilon\xi e^{\mp i\phi}\mp\frac{\Gamma}{2}\omega e^{\mp i\phi}\sin\phi\right) \Biggr],\nonumber\\
	&&f_{21\left(12\right)}=\mp\frac{\tau^2}{4Z}\Biggl[4\xi\left(\omega^2-\varepsilon^2+i\Gamma\omega\right)-2i\Gamma\varepsilon\omega\cos\phi\mp\Biggr.\nonumber\\
	&&\Biggl. ~~~~~~~~~~~~~~~~~~~~~~~~~~~~\mp \Gamma\left(\omega\left(2\omega+i\frac{\Gamma}{2}\right)-\tau^2 \right) \sin\phi \Biggr],\nonumber
	%&&f_{e\left(h\right)22}=-f_{h\left(e\right)11},~f_{h21\left(12\right)}=-f_{e21\left(12\right)},
\end{eqnarray}
where
\begin{eqnarray}
	\label{Zeff}
	&&Z= \frac{\Gamma^2}{8}\tau^4\cos^2\phi -\tau^4\omega\left(2\omega+i\Gamma\right) - \nonumber\\
	&&-2\left(\omega^2-\xi^2\right)\left(\omega^2-\varepsilon^2\right)\left[\left(\omega+i\frac{\Gamma}{2}\right)^2-\varepsilon^2\right]+\\
	&&+\tau^{2}\left(4\omega+i\Gamma\right)\left[\omega\left(\omega^2+i\frac{\Gamma}{2}\omega-\varepsilon^2\right)- i\frac{\Gamma}{2}\varepsilon\xi\cos\phi\right].\nonumber
\end{eqnarray}\\

\textit{4. Overlapping Majorana modes}\\

Next, we analyze transport in the device where two QDs coupled in parallel with the single zero-energy Bogoliubov excitation characterized by the significant overlap of the Majorana wave functions. It is well known that the Kitaev chain with $N$ sites has the same number of lines in the parametric space where the lowest excitation energy becomes zero \cite{hedge-15}, which are defined by the following equations: 
\begin{equation} \label{mu0eq}
\mu=2\sqrt{t^2-\Delta^2}\cos\frac{\pi m}{N+1},~m=1,...,N.
\end{equation}
The Bogoliubov operators of these zero-energy states, $\alpha_{m}$, have form \cite{valkov-21}
\begin{eqnarray} \label{alp0}
&&\alpha_{m} =\frac{1}{S_{m}}\sum\limits_{j=1}^{N}r^{j-1}\sin\frac{\pi mj}{N+1}\cdot\\ &&~~~~~~~~~~~~~~~~~~~~~~~~\cdot\left(a_{j}+a_{j}^{+}+a_{N+1-j}-a_{N+1-j}^{+}\right),\nonumber
\end{eqnarray}
where $S_{m}$ - a normalization factor; $r=\sqrt{\frac{t-\Delta}{t+\Delta}}$. For concreteness, we consider one of such states, $\alpha_{m=1}\equiv\alpha$. As a result, the $u_{j}$ and $v_{j}$ ($j=1,...,N/2$) coefficients are
\begin{eqnarray} \label{uv}
&&u_{j}=u_{N+1-j} =\frac{1}{S}r^{j-1}\left(1+r^{N-j}\right)\sin\frac{\pi j}{N+1},\\ 
&&v_{j}=-v_{N+1-j} =\frac{1}{S}r^{j-1}\left(1-r^{N-j}\right)\sin\frac{\pi j}{N+1}.\nonumber
\end{eqnarray}
Note if $\mu=0$, then the equations (\ref{mu0eq}) and (\ref{uv}) obviously lead to the SKC and $u=v$. Consequently, the Majorana-mode operators are $b_{1}=\alpha^{+}+\alpha=\sum_{j}w_{j}\gamma_{jA}=2u\left(a_{1}^{+}+a_{1}\right)$, $b_{2}=i\left(\alpha^{+}-\alpha\right)=\sum_{j}z_{j}\gamma_{jB}=2iu\left(a_{N}^{+}-a_{N}\right)$. In opposite, if $\mu\neq0$, then the wave functions related to $b_{1}$ and $b_{2}$ overlap. The hybridization is stronger if $t/\Delta\gg1$ resulting in the zero-energy excitation with the bulk spatial distribution.

The inverse Bogoliubov transform gives \cite{tripathi-16}
\begin{equation} \label{a1N}
a_{1}\approx u\alpha + v\alpha^{+},~
a_{N}\approx u\alpha - v\alpha^{+},
\end{equation}
where $u\equiv u_{1}$, $v\equiv v_{1}$. Then, the structure Hamiltonian is
\begin{eqnarray} \label{HDABS}
&&\hat{H}_{D} =\sum\limits_{i=1,2}\xi_{i}d^+_{i}d_{i}-\left(\tau_{1e}d_{1}^{+}+\tau_{2e}d_{2}^{+}\right)\alpha-\\
&&~~~~~~~~~~~~~~~~~~~~~~~~~-\left(\tau_{1h}d_{1}^{+}-\tau_{2h}d_{2}^{+}\right)\alpha^{+}+h.c.,\nonumber
\end{eqnarray}
where $\tau_{je}=\tau_{j}u$, $\tau_{jh}=\tau_{j}v$.

The EOM coefficient matrix and column vector can be written as 
\begin{eqnarray}
\label{Sys_GFABS}
&&\hat{A}_{1}=\left( {\begin{array}{*{10}{c}}
	z_{1e}& Z_{12}&2T_{1}&T_{eh}\\
	-Z_{12}^{*}-2T_{12}&z_{2e}&T_{eh}&-2T_{2}\\
	2T_{1}& T_{eh}&z_{1h}&-Z_{12}^{*}-2T_{12}\\
	T_{eh}&-2T_{2}&Z_{12}&z_{2h}
	\end{array}} \right),\nonumber\\ 
&&\hat{B}_{1}=\left(\omega,0,0,0\right)^{T},
\end{eqnarray}
where $z_{je\left(h\right)}=\omega C_{je\left(h\right)}-\tau_{je}^2-\tau_{jh}^2$, $T_{j}=\tau_{je}\tau_{jh}$, $T_{eh}=\tau_{1h}\tau_{2e}-\tau_{1e}\tau_{2h}$, $T_{12}=\tau_{1e}\tau_{2e}-\tau_{1h}\tau_{2h}$, $Z_{12}=i\Gamma_{12}e^{i\phi}/2-T_{12}$. The EOM system solution in the symmetric transport regime yields
\begin{eqnarray}
\label{aGFABS}
&&f_{11\left(22\right)}=\mp\frac{2\tau_{e}\tau_{h}}{Z}\left[i\Gamma e^{-i\phi}\left(2\tau_{m}\mp \frac{\Gamma}{2}\omega\sin\phi\right)+\right.\nonumber\\
&&\left.~~~~~~~~~~~~~~~~~~~~~~~~~~~+2\omega\left(\omega C-\varepsilon^2\right)-2\tau_{p}\left(\omega+C\right) \right],\nonumber\\
&&f_{21\left(12\right)}=\frac{2\tau_{e}\tau_{h}}{Z}\left[\Gamma\left(\left(\tau_{p}-\frac{\omega}{2}\left(\omega+C\right)\right)\sin\phi-\right.\right.\nonumber\\
&&\left.\left.~~~~~~~~~~~~~~~~~~~~~~~~~~~~~~~~~~~~~~~~~~-i\varepsilon\omega\cos\phi\right)\pm4\tau_{m}\varepsilon \right],\nonumber
%&&f_{h11\left(22\right)}=-f_{e22\left(11\right)},~f_{hji}=-f_{eji},~j\neq i,
\end{eqnarray}
where $C=\omega+i\Gamma/2$, $\tau_{p\left(m\right)}=\tau_{e}^2\pm\tau_{h}^2$.
The GFs' denominator is
\begin{eqnarray}
\label{ZABS}
&&Z=\omega^2\left(\omega^2-\varepsilon^2\right)\left(C^2-\varepsilon^2\right)+\\
&&+2\omega\left[i\frac{\Gamma}{2}\tau_{m}\left(\omega C+\varepsilon^2\right)\cos\phi-\tau_{p}\left(\omega+C\right)\left(\omega C-\varepsilon^2\right)\right]+\nonumber\\
&&+\left[\tau_{p}\left(\omega+C\right)\cos\phi-i\frac{\Gamma}{2}\tau_{m}\right]^2+4\tau_{p}^2\omega C\sin^2\phi-4\tau_{m}^{2}\varepsilon^2.\nonumber
\end{eqnarray}

\section{\label{apxB} The Aharonov-Bohm interferometer containing  noninteracting arrays of zero-energy quantum dots}

\begin{figure}[!h]
	\begin{center}
		\includegraphics[width=0.425\textwidth]{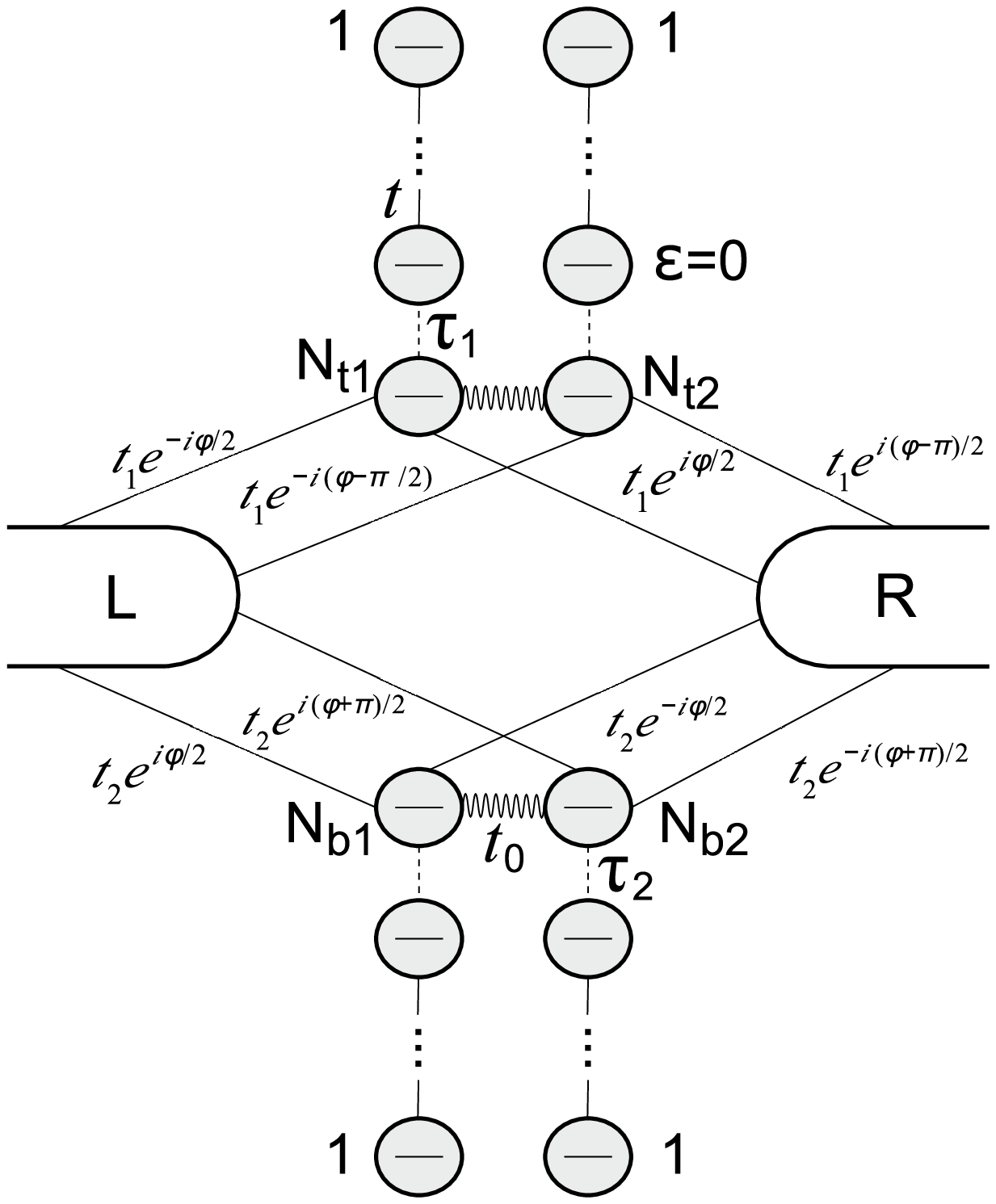}
		\caption{\label{3} The scheme of the Aharonov-Bohm interferometer including top and bottom noninteracting arrays of zero-energy QDs with $N_{t1}+N_{t2}$ and $N_{b1}+N_{b2}$ elements, respectively.}
	\end{center}
\end{figure}

In this Appendix let us examine the AB effect in an interference structure of ordinary zero-energy QDs between left and right normal contacts sketched in Fig. \ref{3}. The necessity to consider such a system is explained by the fact that the interaction between the normal contact and, for example, the SKC implies the coupling with not just the individual Majorana fermions $\gamma_{1A}$ and $\gamma_{NB}$ but also with two Majorana molecules including $\gamma_{1B}$, $\gamma_{2A}$ and $\gamma_{N-1,B}$, $\gamma_{NA}$, respectively. In the situation of the QD-SKC-QD system the separate sleeve consists of two Majorana-fermion chains with two and three constituents. These chains are coupled when $\varepsilon_{j}\neq0$. Thus, if one wants to draw a more explicit analogy between the AB effect in the system of Majorana and fermionic QDs a top (bottom) arm in the latter has to contain two chains with $N_{t1(b1)}$ and $N_{t2(b2)}$ elements of different parity. Additionally, the phase difference of tunneling into these chains is set to $\pi/2$. It allows to show that the behavior (\ref{G1}) takes place if each arm has a single zero-energy eigenstate and the tunneling phase into them differs by $\pi/2$.

\begin{table}
	\caption{\label{t1} The linear-response conductance of the interferometer sketched in Fig. \ref{3} (in the units of $G_{0}$). Parameters: $t,\tau_{1,2},t_{1,2}\neq0$.}
	\begin{indented}\centering
		\lineup
		\item[]\begin{tabular}{ c|c|c }
			\br
			& $N_{t1}$ - even, & $N_{t1}$ - odd \cr
			& $N_{t2}$ - odd & $N_{t2}$ - even  \cr
			\hline 
			$N_{b1}$ - even, & $1,~\phi=\pi n$; & \cellcolor{blue!25} $1,~\phi=\pi \left(n+1/2\right)$;  \cr
			$N_{b2}$ - odd & $0,~\phi\neq\pi n$ & \cellcolor{blue!25}$0,~\phi\neq\pi \left(n+1/2\right)$ \cr
			\hline
			$N_{b1}$ - odd, & \cellcolor{blue!25}$1,~\phi=\pi \left(n+1/2\right)$; & $1,~\phi=\pi n$; \cr
			$N_{b2}$ - even & \cellcolor{blue!25}$0,~\phi\neq\pi \left(n+1/2\right)$ & $0,~\phi\neq\pi n$ \cr
			\br
		\end{tabular}
	\end{indented}
\end{table}

To calculate the conductance the expression (\ref{TLAR1}) can be used where the broadening matrices $\hat{\Gamma}_{e,h}$ have to be replaced by $\hat{\Gamma}_{L,R}$ with the following nonzero blocks:
\begin{eqnarray} \label{GmLR}
&&\hat{\Gamma}_{Ljj}=\Gamma_{j}\hat{\Gamma}_{0},~\hat{\Gamma}_{L12\left(21\right)}=\Gamma_{12}e^{\pm i\phi}\hat{\Gamma}_{0},~\\
&&\hat{\Gamma}_{Rji}=\hat{\Gamma}_{Lji}^{*},~
\hat{\Gamma}_{0}=
\left(\begin{array}{cc}
1  & i \\
-i & 1
\end{array}\right), ~i,j=1,2.\nonumber
\end{eqnarray}
The numerical analysis reveals that the result (\ref{G1}) is valid for any random odd (even) $N_{t1,b2}$ and even (odd) $N_{t2,b1}$ (see the colored cells in Table \ref{t1}). This behavior is not affected by a specific value of $t_{0}$ (which is an analogue of $\varepsilon_{1,2}$ and $\mu$ in the original spinless device). In other words, the problem is effectively reduced to the model studied in \cite{kubala-02} where the real QDs are replaced by two zero-energy eigenstates. For any $t_{0}$ there is the only transport path (not two) in each arm involving these states since the corresponding eigenvectors are proportional to $\sin\pi j/2$, where $j$ - a QD number \cite{guevara-06}. As a result, if the phase of tunneling into the arms is the same (up to the sign), i.e. the parity of $N_{t1}$ and $N_{b1}$ coincide, the conductance peaks occur at $\phi=\pi n$ \cite{kubala-02}. When it differs by $\pi/2$, i.e. $N_{t1}$ and $N_{b1}$ have opposite parity, the corresponding shift appears in the AB effect.

\section*{References}

\bibliography{Majorana}% Produces the bibliography via BibTeX.

\end{document}